\documentclass[11pt,a4paper]{article}
\pdfoutput=1
\usepackage{jheppub-liu}
\usepackage{amsmath,amssymb,graphicx}
\usepackage{float}
\usepackage{subfigure}
\usepackage{mathrsfs}
\usepackage{bm}
\usepackage[all]{xy}
\newcommand{\be}{\begin{equation}}
\newcommand{\bea}{\begin{eqnarray}}
\newcommand{\eea}{\end{eqnarray}}
\newcommand{\ba}{\begin{array}}
\newcommand{\ea}{\end{array}}
\newcommand{\ee}{\end{equation}}
\newcommand{\ml}{\mathcal}

\newcommand{\no}{\nonumber}

\newcommand{\ep}{\epsilon}
\newcommand{\dg}{\dagger}
\newcommand{\de}{\delta}
\newcommand{\al}{\alpha}

\newcommand{\tr}{\mbox{Tr}}

\newcommand{\ihalf}{\frac{i}{2}}
\newcommand{\matr}[2]{\left(\begin{array}{#1}#2\end{array}\right)}

\def\({\left(}
\def\){\right)}
\def\[{\left[}
\def\]{\right]}
\begin{document}

\makeatletter
\gdef\@fpheader{\sc Prepared for submission to JHEP 
}
\makeatother

\title{Integrability of Orbifold ABJM Theories}
\author[a]{Nan Bai}
\author[a]{Hui-Huang Chen}
\author[b]{Xiao-Chen Ding}
\author[a]{De-Sheng Li}
\author[a, c, d, e]{Jun-Bao Wu}

\affiliation[a]{Institute of High Energy Physics, and Theoretical Physics Center for Science Facilities,
Chinese Academy of Sciences, 19B Yuquan Road, Beijing 100049, P.~R.~China}
\affiliation[b]{School of Mathematical Sciences, Capital Normal University, \\105 North Road of West 3rd  Ring, Beijing 100048, P.~R.~China}
\affiliation[c]{School of Physics, Beihang University, Beijing 100191, 37 Xueyuan Road,
P.~R.~China}
\affiliation[d] {Center for High Energy Physics, Peking University, 5 Yiheyuan Rd, Beijing 100871, P.~R.~China}
\affiliation[e]{ University of Chinese Academy of Sciences, 19A Yuquan Road, Beijing 100049, P.~R.~China}

\emailAdd{bainan@ihep.ac.cn}
\emailAdd{chenhh@ihep.ac.cn}
\emailAdd{ingiaohen@163.com}
\emailAdd{lidesheng@ihep.ac.cn}
\emailAdd{wujb@ihep.ac.cn}
\abstract{
Integrable structure has played a very important role in the study of various non-perturbative aspects of planar   Aharony-Bergman-Jafferis-Maldacena (ABJM) theories. In this paper, we showed that this remarkable structure survives after orbifold operation with discrete group $\Gamma(\simeq \mathbb{Z}_n)<SU(4)_R\times U(1)_b$. For general $\Gamma$, we prove the integrability in the scalar sector at the planar two-loop order and get the Bethe ansatz equations (BAEs). The eigenvalues of the anomalous dimension matrix are also obtained. For $\Gamma<SU(4)$, two-loop all-sector and all-loop BAEs are proposed. Supersymmetric orbifolds are  discussed in this framework.}

\arxivnumber{1607.06643}
\maketitle

\section{Introduction}
As a strong-weak duality, AdS/CFT correspondence \cite{Mal97}-\cite{Witten:1998qj} is very powerful in applications which use weakly coupled gravity to study strongly coupled field theory. However, this makes the non-trivial checks of this correspondence very hard since one needs to compute some quantities in the strong coupling limit of field theory to compare with results from the gravity side. Supersymmetric localization \cite{Pestun} and integrability \cite{Beisert:2010jr} are two very important tools to perform such field theoretical computations. These two approaches are complemented by each other. Localization can be utilized  beyond the planar limit but the quantities which it can compute usually should be invariant under the supercharges on which the localization based. When the integrable structure exists, we can compute some quantities which are even non-supersymmetric. However, such theories are quite rare and integrable structure usually only appears in the large N limit.  These two tools also permit us to compute certain quantities at the intermediate values of the coupling constant where neither perturbative gauge theory nor weakly coupled gravity is applicable.

Both four-dimensional ${\cal N}=4$ super Yang-Mills (SYM) theory and three-dimensional  Aharony-Bergman-Jafferis-Maldacena (ABJM) theory \cite{ABJM}  are integrable in the planar limit \cite{Minahan:2002ve}-\cite{c6}.  It is very interesting to see how far one can go by reducing the supersymmetries of the original theory while keeping integrable structure at the same time. For four dimensional case, people have explored a lot through at least three approaches including marginal deformations \cite{Roiban:2003dw}-\cite{Mansson:2008xv}, orbifolding \cite{c4}-\cite{deLeeuw:2012hp} and adding flavors \cite{Chen:2004mu}-\cite{Erler:2005nr}.  Excellent reviews on these results include  \cite{Zoubos:2010kh, vanTongeren:2013gva}. However in three-dimensional case, similar exploration is limited.  In \cite{He:2013hxd}, integrability of planar $\beta$- and $\gamma$-deformed ABJM theories were established at two-loop order in the scalar sector. The anomalous dimension matrices can be expressed as a Hamiltonian acting on an alternative spin chain\footnote{Notice that the $\gamma$-deformation studied in \cite{He:2013hxd} is  different from the one in \cite{Imeroni:2008cr}. The integrability of the latter theory will be discussed in detail in \cite{chenliuwu}.}.   The obtained Hamiltonians have identical form for these theories in the scalar sector, though the former theory has only one deformation parameter, while the latter has three. Comparing with the two-loop scalar-sector Hamiltonian from planar ABJM theory, now in each summand of the Hamiltonian for $\beta$-deformed ABJM theory, the next-to-nearest permutation term attains a certain phase depending on the charges of the three involved sites under two global $U(1)$'s which are used to perform the $\beta$-deformations. To obtain the needed transfer matrices, we need to deform the four R-matrices by similar phase factors to satisfy Yang-Baxter equations and produce the wanted Hamiltonian at the same time. This deformation is of Drinfeld-Reshetikhin form. A double scaling limit of $\gamma$-deformed ABJM theory was considered in \cite{CGK} which leads to an integrable theory of interacting fermions and scalars following four-dimensional consideration in \cite{Gurdogan:2015csr} (some subtleties of this limit were also studied in \cite{Sieg:2016vap}). This showed that integrable Chern-Simons-matter theories with less supersymmetry can have new interesting feature.
And as in four dimensional case \cite{Zoubos:2010kh}, in $\beta$/$\gamma$-deformed and orbifold ABJM theories, states with single magnon can be physical and detailed study on them may be simpler in many aspects than on the excited states in ABJM theory where at least two magnons are needed.

In this paper, we will focus on integrability of planar orbifold ABJM theories. Orbifolding is a widely used technique to obtain gauge theories from a parent one \cite{Kachru:1998ys, Bershadsky:1998mb, Bershadsky:1998cb}. It is carried out by starting with a discrete subgroup of the global symmetry group of the original theory. One can get various quiver gauge theories with less supersymmetry based on different discrete subgroups when the former one is supersymmetric. One of the advantages of the orbifolding operation
is that the obtained  theories inherit some good properties of the parent theory. In this paper, the parent theory is the ABJM  theory \cite{ABJM} which is the low energy effective theory on the worldvolume of N coincident M2-branes at the  $\mathbb{C}^4/\mathbb{Z}_k$ orbifold singularity. The global (super)symmetry of ABJM theory is $OSp(6|4)\times U(1)_b$, direct product of a simple supergroup and a $U(1)$ factor.
This is distinct from the global symmetry of $\ml{N}=4$ SYM which is just a simple supergroup $PSU(2, 2|4)$. In \cite{c1}, two concrete quiver gauge theories with the residual $\mathcal{N}=4$ supersymmetries (non-chiral  orbifold) and $\mathcal{N}=2$ supersymmetries (chiral orbifold) have been established through two different $\mathbb{Z}_n$ orbifoldings in ABJM field theory. Other orbifold ABJM theories are discussed in \cite{Terashima:2008ba, Imamura:2008ji, Berenstein}. In this paper, we only consider the case that $\Gamma$ is isomorphic to $\mathbb{Z}_n$.

We start with planar two-loop order and focus on the scalar sector which is closed at this order. We consider the generic case with $\Gamma<SU(4)_R\times U(1)_b$. The composite operators of the
orbifold theory can be expressed compactly using the fields in the parent ABJM theory with twist matrix inserted in the trace and with the projection condition imposed on the fields in the parent theory .
A straightforward computation shows that only two terms of the Hamiltonian were twisted by some phase factors whose precise forms depend on the charges of the involved sites under the action of $\Gamma$. To get transfer matrices which can produce this new Hamiltonian, we only  need to insert certain constant matrices which act on the auxiliary spaces inside the traces. One can demonstrate that choosing the inserted matrices to be  diagonal  will make the RTT relations hold. By suitable choices of such matrices, we can produce the desired Hamiltonian. This completes the proof the integrability of general orbifold ABJM theories at planar two loop order in the scalar sector. Using algebraic Bethe ansatz, we find the Bethe ansatz equations (BAEs) in this sector at two-loop and give the constraints from the trace property and twist condition. The eigenvalues of the anomalous dimension matrix (ADM) are expressed using the Bethe roots.

Then we concentrate on the case with $\Gamma<SU(4)_R$ and generalize the above results to proposals for all-sector and all-loop order.
The leading-order all-sector results can be employed based on the prescription of Beisert-Roiban \cite{c2} after obtaining the charges for each simple root of the superalgebra and the vacuum. The all-order asymptotic results are obtained similarly based on all-loop asymptotic BAEs for planar ABJM theory \cite{c9}. As non-trivial consistency checks, we show that the BAEs we obtained satisfy both the fermionic duality and dynamic duality conditions. Finally we analyse the condition on the charges for the orbifolding to preserve $\ml{N}=2$ and $\ml{N}=4$ supersymmetries. We also confirm these results by using the fact that the orbifold ABJM theory is the low energy effective theory of $N$ membranes placed at the orbifold singularity $\mathbb{C}^4/(\Gamma\times \mathbb{Z}_{|\Gamma|k})$ \cite{Berenstein}.

The remaining part of this paper is organized as follows, in the next section and section~\ref{section3} we study in detail the integrability of orbifold ABJM theories in the scalar sector at two loop level. In sections~\ref{section4} and \ref{section5}, we will obtain the two-loop all-sector and all-loop results. Finally, we will discuss the supersymmetric orbifold theories. Some technical details will be put in three appendices.

\section{Two-loop Hamiltonian from orbifold ABJM theories}
 As mentioned in the introduction, in this section, we will consider orbifold based on group $\Gamma(\simeq \mathbb{Z}_n)< SU(4)_R\times U(1)_b$ and focus on the scalar sector which is closed at two-loop order.
\subsection{Basic Ingredients of Orbifolding in Gauge Theory}
Now we will set up some necessary knowledge of orbifold gauge theory and our notation will follow that of \cite{c2,c3} closely. 
We consider to perform orbifolding using discrete subgroup $\Gamma\simeq \mathbb{Z}_n$ of $SU(4)_R\times U(1)_b$ which means to start with ABJM theory with gauge group $U(nN)\times U(nN)$ and impose the following projection condition on gauge fields and the scalar fields
\bea
\gamma(g) A \gamma^{-1}(g)&=&A,\\
\gamma(g)\hat{A} \gamma^{-1}(g)&=&\hat{A},\\
\gamma(g)\left(R(g)^I_J Y^J\right)\gamma^{-1}(g)&=&Y^I,
\eea
where $R(g)$ is a matrix representation of $\Gamma$ acting on the indices $I,J=1 \cdots 4$ of $Y^I$ and $\gamma(g)$ is acting on the color space with the color indices suppressed.
The projection condition on fermions is
\be \gamma(g)\left(R^\prime(g)^I_J\bar{\psi}^J\right)\gamma^{-1}(g)=\bar{\psi}^I. \ee
Notice that when $g=(g_1, g_2)\in SU(4)_R\times U(1)_b$, we have $R(g)=R(g_1)R(g_2)$ and $R^\prime(g)=R^\prime(g_1)R^\prime(g_2)=R(g_1)R(g_2^{-1})$, since $\bar{\psi}^I$ and $Y^I$
have opposite $U(1)_b$ charges.
The resulting theory is a quiver theory with gauge group $U(N)^{2n}$. If the element $g$ is the generator of $\mathbb{Z}_n$, the matrix representation $\gamma(g)$ will have the form
\bea
\gamma(g)=\mbox{diag}\left(\mathbb{I}_{N\times N}, \omega \mathbb{I}_{N\times N},\cdots \omega^{n-1}\mathbb{I}_{N\times N}\right),\qquad \omega=e^{\frac{2\pi i}{n}}.
\eea
For the sake of simplicity, we also require that the field $Y^I$ has definite $\Gamma(<SU(4)_R\times U(1)_b)$ charge, then $R(g)$ will take the diagonal form $R(g)^I_J=\delta^I_J \omega^{s_I}$ and the constraint on  the field $Y^I$ becomes
\bea
Y^I=\omega^{s_I}\gamma Y^I \gamma^{-1}. \label{orbicon}
\eea
Here and the following by $\gamma$, we always mean $\gamma(g)$.
\par By orbifolding, the field $Y^I$ can be viewed as a $n \times n$ matrix with elements also being $N \times N$ matrices and only some components will survive due to the condition (\ref{orbicon}). Then the orbifold theory can be formulated in terms of those decomposed fields however the action turns out to be quite complicated \cite{c1,c4}. In our paper, we will use the field $Y^I$ in the parent theory and focus on the following single trace operators,
\bea
\tr \left(\gamma^m Y^{I_1}Y^{\dg}_{J_1}\cdots Y^{I_L} Y^{\dg}_{J_L}\right),\quad  m=0,1,\cdots n-1,\quad L\geq 2. \label{obasis}
\eea
Operators with the same $m$ constitute the $m$-th twisted sector and $m=0$ corresponds to the untwisted sector. If we move one $\gamma$ to pass all the fields behind and use the cyclic property of the trace to move it back, we find an overall phase factor appear. The composite operators will have the possibility to be non-vanishing only when this phase factor is trivial, hence lead to the twist constraint
\bea
\frac{1}{n}\left(-\sum_{k}^Ls_{I_k}+\sum_{k}^Ls_{J_k}\right) \in \mathbb{Z}.
\eea
Furthermore, this local operator can also be seen as a closed alternating spin chain state
\bea
|\ml{O}\rangle=|\gamma^m;I_1,\bar{J}_1,\cdots,I_{L},\bar{J}_L\rangle.
\eea
\subsection{Anomalous Dimension Matrix of Composite Operators in Twisted Sector}
We now find the anomalous dimensions for these gauge invariant scalar operators. An important fact is that the operators belonging to different twisted sectors do not mix with each other. Thus in the following discussions we will stay in a fixed $m$-th twisted sector. Before any further computations, let us recall that for parent ABJM theory, in the planar limit and at 2-loop order, the anomalous dimension matrix $\Gamma$ consists of local Hamiltonian of three adjacent sites \cite{c5,c6},
\bea
\Gamma=\frac{\lambda^2}{2}\sum^{2L}_{i=1}\left(2-2P_{i,i+2}+P_{i,i+2}K_{i,i+1}+K_{i,i+1}P_{i,i+2}\right)=\frac{\lambda^2}{2}\sum^{2L}_{i=1} H_{i,i+1,i+2}.
\eea
where $\lambda={N}/{k}$ and $P$, $K$ are the permutation and trace operators acting on the tensor product of two vector spaces defined as
\bea
P_{i_1,i_2}^{j_1,j_2}=\delta^{j_2}_{i_1}\delta^{j_1}_{i_2},\quad K_{i_1,j_2}^{j_1,i_2}=\delta^{i_2}_{i_1}\delta^{j_1}_{j_2}.
\eea
For the orbifold ABJM theories, the anomalous dimension matrix is obtained by expressing local interaction terms $\mathcal{H}$ of ABJM theory in the operator basis (\ref{obasis}). If $\gamma^m$ do not appear in the interaction region, we get the same local Hamiltonian as the parent ABJM theory,
\bea
\mathcal{H}\circ Y^{I_i}Y^{\dg}_{I_{i+1}}Y^{I_{i+2}}=\left(H_{i,i+1,i+2}\right)^{I_i,J_{i+1},I_{i+2}}_{J_{i},I_{i+1},J_{i+2}}Y^{J_i}Y^{\dg}_{J_{i+1}}Y^{J_{i+2}}.
\eea
If $\gamma^m$ is present in the interaction region, we should move it away either to the left or to the right for the convenience of solving the problem. In our case, the non-trivial interactions only reside in the first and the last two sites of the spin chain and the modified local Hamiltonian are derived as follows. For the interactions among the $(2L-1)$-th,the $2L$-th and the 1st site, on one hand, we have,
\bea
{\mathcal{H}}_{2L-1,2L,1}\circ Y^{I_{2L-1}}Y^{\dg}_{I_{2L}}\gamma^m Y^{I_1}=({H}^{orbi}_{2L-1,2L,1})^{I_{2L-1},J_{2L},I_1}_{J_{2L-1},I_{2L},J_1}Y^{J_{2L-1}}Y^{\dg}_{J_{2L}}\gamma^m Y^{J_1}.
\eea
where ${H}^{orbi}$ represents the orbifold Hamiltonian. On the other hand, we also have
\bea
{\mathcal{H}}_{2L-1,2L,1}\circ Y^{I_{2L-1}}Y^{\dg}_{I_{2L}}\gamma^m Y^{I_1}&=&\omega^{-ms_{I_1}}{\mathcal{H}}_{2L-1,2L,1}\circ Y^{I_{2L-1}}Y^{\dg}_{I_{2L}} Y^{I_1} \gamma^m\\\no
&=&\omega^{-ms_{I_1}}({H}_{2L-1,2L,1})^{I_{2L-1},J_{2L},I_1}_{J_{2L-1},I_{2L},J_1}Y^{J_{2L-1}}Y^{\dg}_{J_{2L}} Y^{J_1} \gamma^m\\\no
&=&\omega^{-ms_{I_1}+ms_{J_1}}({H}_{2L-1,2L,1})^{I_{2L-1},J_{2L},I_1}_{J_{2L-1},I_{2L},J_1}Y^{J_{2L-1}}Y^{\dg}_{J_{2L}}\gamma^m Y^{J_1},
\eea
where we have used the relation $\gamma^m Y^I=\omega^{-ms_I}Y^I\gamma^m$ deduced from (\ref{orbicon}). Finally we get
\bea
({H}^{orbi}_{2L-1,2L,1})^{I_{2L-1},J_{2L},I_1}_{J_{2L-1},I_{2L},J_1}=\omega^{-ms_{I_1}+ms_{J_1}}({H}_{2L-1,2L,1})^{I_{2L-1},J_{2L},I_1}_{J_{2L-1},I_{2L},J_1}. \label{tbc1}
\eea
Similarly, when acting on the $2L$-th ,the 1st and the 2nd sites, we shift the generator $\gamma^m$ to the left side and obtain
\bea
({H}^{orbi}_{2L,1,2})^{J_{2L},I_{1},J_2}_{I_{2L},J_{1},I_2}=\omega^{-ms_{I_{2L}}+ms_{J_{2L}}}({H}_{2L,1,2})^{J_{2L},I_{1},J_2}_{I_{2L},J_{1},I_2}. \label{tbc2}
\eea
Therefore, the orbifold ABJM Hamiltonian reads
\bea
H^{orbi}=\frac{\lambda^2}{2}\sum^{2L-2}_{i=1} H_{i,i+1,i+2}+\frac{\lambda^2}{2}\left({H}^{orbi}_{2L-1,2L,1}+{H}^{orbi}_{2L,1,2}\right). \label{twistbc}
\eea

\section{Algebraic Bethe Ansatz of the Orbifold ABJM Model}\label{section3}
The Hamiltonian derived above can be seen as a spin chain Hamiltonian with twisted boundary conditions. In this section we will give an explicit construction to show the integrability of this model and compute  the eigenvalues of the Hamiltonian.
\subsection{Integrability of Orbifold ABJM Hamiltonian}
In order to demonstrate the integrability, the starting object is the R-matrix which satisfy the Yang-Baxter equation (YBE). For the orbifold ABJM theories, we use the same four R-matrices as those defined in the case of period spin chain \cite{c5,c6},
\bea
R_{ab}(u)=u-P_{ab}:\qquad V_a\otimes V_b \rightarrow V_a\otimes V_b,\\
R_{\bar{a}\bar{b}}(u)=u-P_{\bar{a}\bar{b}}:\qquad V_{\bar{a}}\otimes V_{\bar{b}} \rightarrow V_{\bar{a}}\otimes V_{\bar{b}},\\
R_{a\bar{b}}(u)=u+K_{a\bar{b}}:\qquad V_a\otimes V_{\bar{b}} \rightarrow V_a\otimes V_{\bar{b}},\\
R_{\bar{a}b}(u)=u+K_{\bar{a}b}:\qquad V_{\bar{a}}\otimes V_b \rightarrow V_{\bar{a}}\otimes V_b.
\eea
where $V_i$ and $V_{\bar{i}}$ denote the fundamental and anti-fundamental representation space of $SU(4)$ respectively. The R-matrices satisfy the following six YBEs \cite{c5,c6},
\bea
R_{ab}(u-v)R_{ac}(u)R_{bc}(v)&=&R_{bc}(v)R_{ac}(u)R_{ab}(u-v),\\
R_{ab}(u-v)R_{a\bar{c}}(u)R_{b\bar{c}}(v)&=&R_{b\bar{c}}(v)R_{a\bar{c}}(u)R_{ab}(u-v),\\
R_{\bar{a}\bar{b}}(u-v)R_{\bar{a}c}(u)R_{\bar{b}c}(v)&=&R_{\bar{b}c}(v)R_{\bar{a}c}(u)R_{\bar{a}\bar{b}}(u-v),\\
R_{\bar{a}\bar{b}}(u-v)R_{\bar{a}\bar{c}}(u)R_{\bar{b}\bar{c}}(v)&=&R_{\bar{b}\bar{c}}(v)R_{\bar{a}\bar{c}}(u)R_{\bar{a}\bar{b}}(u-v),\\
R_{a\bar{b}}(u-v-2)R_{ac}(u)R_{\bar{b}c}(v-2)&=&R_{\bar{b}c}(v-2)R_{ac}(u)R_{a\bar{b}}(u-v-2),\\
R_{a\bar{b}}(u-v-2)R_{a\bar{c}}(u-2)R_{\bar{b}\bar{c}}(v)&=&R_{\bar{b}\bar{c}}(v)R_{a\bar{c}}(u-2)R_{a\bar{b}}(u-v-2).
\eea
By the standard procedure, the next step is to construct the monodromy matrices using these R-matrices, we have
\bea
T_0(u)=M_0 R_{01}(u)R_{0\bar{1}}(u-2)R_{02}(u)R_{0\bar{2}}(u-2)\cdots R_{0L}(u)R_{0\bar{L}}(u-2), \label{mono1}\\
T_{\bar{0}}(u)=\bar{M}_{\bar{0}} R_{{\bar{0}}1}(u-2)R_{{\bar{0}}\bar{1}}(u)R_{{\bar{0}}2}(u-2)R_{{\bar{0}}\bar{2}}(u)\cdots R_{{\bar{0}}L}(u-2)R_{{\bar{0}}\bar{L}}(u). \label{mono2}
\eea
where $0$ and $\bar{0}$ refer to auxiliary spaces in the SU(4) fundamental and anti-fundamental representations respectively. Comparing with the T-matrices for the periodic spin chain, we modify them by inserting two additional matrices $M$ and $\bar{M}$ in the auxiliary spaces $V_0$ and $V_{\bar{0}}$ so that they can generate the twisted boundary terms in equation (\ref{twistbc}) \cite{Zabrodin:2007rq}. The precise form of these two matrices will be determined later by demanding that the obtained Hamiltonian is the same as the one from the orbifold ABJM theories (up to an overall constant factor and shifting by term proportional to identity operator). Here we first show that when $M$ and $\bar{M}$ are diagonal and $M\bar{M}$ is proportional to identity matrix, the obtained Hamiltonian is integrable. In this case it is easy to show that
\bea
\left[R_{ab}(u), M_a M_b\right]=0,\\
\left[R_{a\bar{b}}(u), M_a \bar{M}_{\bar{b}}\right]=0,\\
\left[R_{\bar{a}\bar{b}}(u), \bar{M}_{\bar{a}}\bar{M}_{\bar{b}}\right]=0.
\eea
where the  indices of $M$ and $\bar{M}$ denote on which site they act. Therefore we have the following important equations known as the RTT relations in the literature,
\bea
R_{ab}(u-v)T_a(u)T_b(v)&=&T_b(v)T_a(u)R_{ab}(u-v), \label{fcr1}\\
R_{\bar{a}\bar{b}}(u-v)T_{\bar{a}}(u)T_{\bar{b}}(v)&=&T_{\bar{b}}(v)T_{\bar{a}}(u)R_{\bar{a}\bar{b}}(u-v),\\
R_{a\bar{b}}(u-v-2)T_a(u)T_{\bar{b}}(v)&=&T_{\bar{b}}(v)T_a(u)R_{a\bar{b}}(u-v-2).
\eea
By tracing over the auxiliary spaces of monodromy T-matrices, we obtain the transfer matrices
\bea
\tau(u)=\tr_0 T_0(u),\qquad \bar{\tau}(u)=\tr_{\bar{0}} T_{\bar{0}}(u).
\eea
Then the above RTT relations lead to
\bea
\left[\tau(u),\tau(v)\right]=0,\\
\left[\bar{\tau}(u),\bar{\tau}(v)\right]=0,\\
\left[\tau(u),\bar{\tau}(v)\right]=0.
\eea
for arbitrary $u$ and $v$. Expanding $\tau(u)$ and $\bar{\tau}(u)$ in terms of $u$, we find that the coefficients are mutually commuting  and can be seen as the conserved charges. Of our interests is a certain combination of these conserved quantities given below because they will correspond to the Hamiltonians of our system,
\bea
&&H_{1}=\tau(u)^{-1}\frac{d}{du}\tau(u)\bigg|_{u=0},\\
&&H_{2}=\bar{\tau}(u)^{-1}\frac{d}{du}\bar{\tau}(u)\bigg|_{u=0}.
\eea
After some computations ,we find
\bea
H&=&H_1+H_2\\\no
&=&\frac{1}{2}\sum_{i=1}^{2L-2}\left(-2-2P_{i,i+2}+P_{i,i+2}K_{i,i+1}+K_{i,i+1}P_{i,i+2}\right)\\\no
&-&1-M^{-1}_1P_{2L-1,1}M_1+\frac{1}{2}K_{2L-1,2L}M^{-1}_1K_{2L,1}M_1+\frac{1}{2}M^{-1}_1K_{2L,1}M_1K_{2L-1,2L}\\\no
&-&1-P_{2L,2}\bar{M}^{-1}_{2L}\bar{M}_2+\frac{1}{2}P_{2L,2}\bar{M}^{-1}_{2L}\bar{M}_2 K_{2,1} + \frac{1}{2}K_{2,1} P_{2L,2}\bar{M}^{-1}_{2L}\bar{M}_2. \label{totalH}
\eea
The details of the computations are put in the Appendix~\ref{appendixa}. We would like to know the component forms of the boundary terms of the above Hamiltonian and for this purpose we first clarify our convention for the matrix indices as follows\footnote{We only demonstrate this convention for case when
all indices are in the $\bf{4}$ representation. The convention for other cases is similar.},
\bea
(AB)^{i_1,i_2}_{j_1,j_2}=(A)^{a,\,\,b}_{j_1,j_2}(B)^{i_1,i_2}_{a,\,\,b}.
\eea
Hence we have
\bea
(M^{-1}_1 P_{1,2L-1} M_1)^{I_{2L-1},I_1}_{J_{2L-1},J_1}&=&(M^{-1}_1)^b_{J_1}(P_{1,2L-1})^{a,I_{2L-1}}_{b,J_{2L-1}}(M_1)^{I_1}_a\\\no
&=&m_{I_1}\de^{I_1}_a\cdot \de^a_{J_{2L-1}}\de^{I_{2L-1}}_b\cdot m^{-1}_{J_1}\de^b_{J_1}\\\no
&=&m_{I_1} m^{-1}_{J_1}(P_{1,2L-1})^{I_{2L-1},I_1}_{J_{2L-1},J_1}.
\eea
\bea
(K_{2L-1,2L} M^{-1}_1 K_{2L,1} M_1)^{I_{2L-1},J_{2L},I_1}_{J_{2L-1},I_{2L},J_1}&=&(K_{2L-1,2L})^{J_{2L},I_{2L-1}}_{\,\,a,\,J_{2L-1}}(M^{-1})^c_{J_1}(K_{2L,1})^{b,a}_{c,I_{2L}}(M)_b^{I_1}\\\no
&=&(K_{2L-1,2L})^{J_{2L},I_{2L-1}}_{\,\,a,\,J_{2L-1}}m_{I_1} m^{-1}_{J_1}\delta^a_{J_1}\delta^{I_1}_{I_{2L}}\\\no
&=&m_{I_1} m^{-1}_{J_1}(K_{2L-1,2L}  K_{2L,1} )^{I_{2L-1},J_{2L},I_1}_{J_{2L-1},I_{2L},J_1}.
\eea
\bea
(P_{2L,2}\bar{M}^{-1}_{2L}\bar{M}_2)^{J_{2L},J_2}_{I_{2L},I_2}&=&(P_{2L,2})^{J_{2L},J_2}_{\,\,b,\,\,\,a}(\bar{M}^{-1}_{2L})^{b}_{I_{2L}}(\bar{M}_2)^{a}_{I_2} \\\no
&=&\de^{J_{2L}}_a\de^{J_2}_b\bar{m}^{-1}_{I_{2L}}\de^b_{I_{2L}}\bar{m}_{I_2}\de^a_{I_2}\\\no
&=&\bar{m}^{-1}_{I_{2L}}\bar{m}_{I_2}(P_{2L,2})^{J_{2L},J_2}_{I_{2L},I_2},
\eea
where $m_i$ and $\bar{m}_i$ $i=1,2,3,4$ are the diagonal elements of $M$ and $\bar{M}$. Comparing these results with the equations (\ref{tbc1}) and (\ref{tbc2}), one can fix the matrices $M$ and $\bar{M}$ as
\bea
M&=&\mbox{diag}\,(\omega^{-ms_{1}},\omega^{-ms_{2}},\omega^{-ms_{3}},\omega^{-ms_{4}}),\\
\bar{M}&=&\mbox{diag}\,(\omega^{ms_{1}},\omega^{ms_{2}},\omega^{ms_{3}},\omega^{ms_{4}}).
\eea
Notice that $M\bar{M}$ is identity matrix which, with the fact that $M$ and $\bar{M}$ are diagonal,  guarantees the integrability as mentioned above.
So our conclusion is that by inserting the above two diagonal matrices into the monodromy matrices, we derived a Hamiltonian nearly the same as the one obtained in the field theory side only up to a shift of $3L \mathbb{I}$ and an overall factor $\lambda^2$ which do not affect the integrability of the model. This completes the proof of the integrability of planar orbifold ABJM theories in the scalar sector at the two-loop order.

\subsection{Eigenvalues of Spin Chain Hamiltonian and Bethe Ansatz Equations}
In this section we consider the diagonalisation of the corresponding transfer matrices. In the seminal paper \cite{c7}, the eigenstates of the Hamiltonian for a very general inhomogeneous spin chain with different spin on each site were constructed by means of the nested algebraic Bethe ansatz method. We find the related results can also apply to our alternating spin chain with twisted boundary conditions. However, here, we will use a much simpler method to obtain the Bethe ansatz equations.\footnote{Such treatment for $SU(N)$ spin chain can be found in the lecture notes by N.~Beisert \cite{c8}.} First we select the ground state as
\bea
|\Omega\rangle=|\gamma^m; 1\bar{4}\cdots 1\bar{4}\rangle.
\eea
which corresponds to the chiral primary operator $\tr (\gamma^m(Y^1Y^{\dg}_4)^L)$. Then we write the monodromy matrix as
\bea
T_0=\left(
\ba{cccc}
T_1&B_1&*&*\\C_1&T_2&B_2&*\\{*}&C_2&T_3&B_3\\{*}&{*}&C_3&T_4
\ea
\right)
\eea
For this selected vacuum, we find the three super-diagonal elements $B_1=T^1_2$,$B_2=T^2_3$,$B_3=T^3_4$ serve as the creation operators while the other three sub-diagonal ones $C_1=T^2_1$,$C_2=T^3_2$,$C_3=T^4_3$ as the annihilation operators. They also correspond to the simple roots of $SU(4)$ Lie algebra.
\par The excited states can be constructed by acting three kinds of creation operators on the vacuum state,
\bea
\prod_{k=1}^{K_r}B_2(u_{2k})\prod_{j=1}^{K_u}B_1(u_{1j})\prod_{n=1}^{K_v}B_3(u_{3n})|\Omega\rangle, \label{exs}
\eea
where $u_{1j}=i u_j+1/2,u_{2k}=ir_k+1,u_{3n}=iv_n+3/2$ with $1\leq j\leq K_u, 1\leq k\leq K_r, 1\leq n\leq K_v$ are three sets of Bethe roots.
Then the eigenvalue of the transfer matrix $\tau(u)$ can be found by using the commutation relations between $T_i,i=1,2,3,4$ and $B_i,i=1,2,3$ originated from the eq.~(\ref{fcr1}) by throwing the unwanted terms,
\bea
\Lambda(u)&=&\omega^{-ms_1}(u-1)^L(u-2)^L\prod_{j=1}^{K_u}\frac{u-i u_j+\frac{1}{2}}{u-i u_j-\frac{1}{2}}\\\no
&+&\omega^{-ms_2}u^L(u-2)^L\prod_{j=1}^{K_u}\frac{u-i u_j-\frac{3}{2}}{u-i u_j-\frac{1}{2}}\prod_{k=1}^{K_r}\frac{u-i r_k}{u-i r_k-1}\\\no
&+&\omega^{-ms_3}u^L(u-2)^L\prod_{n=1}^{K_v}\frac{u-i v_n-\frac{1}{2}}{u-i v_n-\frac{3}{2}}\prod_{k=1}^{K_r}\frac{u-i r_k-2}{u-i r_k-1}+\omega^{-ms_4}u^L(u-1)^L\prod_{n=1}^{K_v}\frac{u-i v_n-\frac{5}{2}}{u-i v_n-\frac{3}{2}}.
\eea
For the eigenvalue of $\bar{\tau}(u)$, it can be found from the conjugation condition $\bar{\Lambda}(u)=\Lambda(2-u^*)^*$ \cite{c5}
\bea
 \bar{\Lambda}(u)&=&\omega^{ms_1}u^L(u-1)^L\prod_{j=1}^{K_u}\frac{u-i u_j-\frac{5}{2}}{u-i u_j-\frac{3}{2}}\\\no
&+&\omega^{ms_2}u^L(u-2)^L\prod_{j=1}^{K_u}\frac{u-i u_j-\frac{1}{2}}{u-i u_j-\frac{3}{2}}\prod_{k=1}^{K_r}\frac{u-i r_k-2}{u-i r_k-1}\\\no
&+&\omega^{ms_3}u^L(u-2)^L\prod_{n=1}^{K_v}\frac{u-i v_n-\frac{3}{2}}{u-i v_n-\frac{1}{2}}\prod_{k=1}^{K_r}\frac{u-i r_k}{u-i r_k-1}+\omega^{ms_4}(u-2)^L(u-1)^L\prod_{n=1}^{K_v}\frac{u-i v_n+\frac{1}{2}}{u-i v_n-\frac{1}{2}}.
\eea
The Bethe ansatz equations (BAEs) can be readily obtained by demanding that the residue vanishes at each potential pole of $\Lambda(u)$,
\bea
\omega^{-ms_1+ms_2}\left(\frac{u_j+i/2}{u_j-i/2}\right)^L=\prod_{k\neq j}^{K_u}\frac{u_j-u_k+i}{u_j-u_k-i}\prod_{k=1}^{K_r}\frac{u_j-r_k-i/2}{u_j-r_k+i/2},\label{BAE1}\\
\omega^{-ms_2+ms_3}=\prod_{k\neq j}^{K_r}\frac{r_j-r_k+i}{r_j-r_k-i}\prod_{k=1}^{K_u}\frac{r_j-u_k-i/2}{r_j-u_k+i/2}\prod_{k=1}^{K_v}\frac{r_j-v_k-i/2}{r_j-v_k+i/2},\label{BAE2}\\
\omega^{-ms_3+ms_4}\left(\frac{v_j+i/2}{v_j-i/2}\right)^L=\prod_{k\neq j}^{K_v}\frac{v_j-v_k+i}{v_j-v_k-i}\prod_{k=1}^{K_r}\frac{v_j-r_k-i/2}{v_j-r_k+i/2}.\label{BAE3}
\eea
The consistency of the theory guarantees that we could get the same sets of Bethe ansatz equations from $\bar{\Lambda}(u)$ instead, as one can easily check.
\par Now let us investigate the twist constraint for the excited state which is largely due to an implicit charge conservation condition. Note that the component of monodromy matrix T is
\bea
&&\left(T_0(u)\right)^{b;i_1,j_2,\cdots,i_{2L-1},j_{2L}}_{a;j_1,i_2,\cdots,j_{2L-1},i_{2L}}\\\no
&=&\left(M_0\right)^{c_1}_{a}\left(R_{01}(u)\right)^{c_2,i_1}_{c_1,j_1}\left(R_{02}(u-2)\right)^{c_3,j_2}_{c_2,i_2}\cdots\left(R_{0,2L-1}(u)\right)^{c_{2L},i_{2L-1}}_{c_{2L-1},j_{2L-1}}\left(R_{0,2L}(u-2)\right)^{b,j_{2L}}_{c_{2L},i_{2L}},
\eea
where $a,b,c_n,n=1,\cdots,2L$ represent the indices of the auxiliary space and $i_n,j_n,n=1,\cdots,2L$ are the indices of quantum spaces. If allocating each index $i$ of $V_i$ a phase $s_i$ and $i'$ of $\bar{V}_{i'}$ a phase $\bar{s}_{i'}$ with obvious relation $\bar{s}_{i'}=-s_{i'}$, we find the total phases are conserved under the action of three braiding operators $I,P$ and $K$,
\bea
(I)^{k,l}_{i,j}=\de^k_i\de^l_j\quad &\rightarrow& \quad s_k+s_l=s_i+s_j,\\
(P)^{k,l}_{i,j}=\de^k_j\de^l_i\quad &\rightarrow& \quad s_k+s_l=s_j+s_i,\\
(K)^{k,j}_{i,l}=\de^k_l\de^j_i\quad &\rightarrow& \quad s_k+\bar{s}_l=s_k-s_l=0,\\\no
&&\quad s_i+\bar{s}_j=s_i-s_j=0.
\eea
Since the building blocks of the monodromy matrix are R-matrices which entirely consists of these three operators and $M$ is diagonal, the whole process obey the phase conservation law,
\bea
s_b+\sum_{k=1}^{2L}s_{i_k}=s_a+\sum_{k=1}^{2L}s_{j_{k}}.
\eea
So the net phase of the quantum space is $s_a-s_b$ under the action of $T^b_a$. Note that the phase of vacuum state is $L(s_1-s_4)$, then the phase of the excited state (\ref{exs}) become
\be\label{twistconstraint}
K_u(s_2-s_1)+K_r(s_3-s_2)+K_v(s_4-s_3)+L(s_1-s_4).
\ee
Therefore the twist constraint turn out to be
\bea
\frac{1}{n}\left(K_u(s_2-s_1)+K_r(s_3-s_2)+K_v(s_4-s_3)+L(s_1-s_4)\right) \in \mathbb{Z}.
\eea
\par The shift operator and the corresponding total momentum are defined as
\bea
\Pi=e^{2iP}=\frac{1}{2^{2L}}\tau(0)\bar{\tau}(0).
\eea
In the Appendix~\ref{appendixb} we will show that the shift operator acts trivially on physical state. Now given the eigenvalues above, we find
\be\label{ZMC}
1=\frac{1}{2^{2L}}\Lambda(0)\bar{\Lambda}(0)=\omega^{m(s_4-s_1)}
\prod_{i=1}^{K_u}\frac{u_i+\frac{i}{2}}{u_i-\frac{i}{2}}\prod_{j=1}^{K_v}\frac{v_j+\frac{i}{2}}{v_j-\frac{i}{2}},
\ee
which is the zero momentum condition for the twisted spin chain. As mentioned above, by a shift of $3L$ and then multiplied by $\lambda^2$, we find the energy of the spin chain which is dual to the anomalous dimension $\gamma$ of the orbifold ABJM theories,
\bea
E=\lambda^2\left(3L+\frac{d}{du}\log (\Lambda(u)\bar{\Lambda}(u))|_{u=0}\right)=\lambda^2\left(\sum_{j=1}^{K_u}\frac{1}{u_j^2+\frac{1}{4}}+\sum_{j=1}^{K_v}\frac{1}{v_j^2+\frac{1}{4}}\right).
\eea
\section{Orbifold Bethe Ansatz}\label{section4}
Having obtained the orbifold Bethe equations for $SU(4)$ sector, now we go toward all-sector results. From now on, we will restrict to the case
with $\Gamma< SU(4)_R$.
The leading order\footnote{For Chern-Simons-matter theories, leading order means two loop level.} Bethe ansatz equations for ABJM theory read \cite{c5, c6},
\bea\label{LOBAE}
\left(\frac{u_{j,k}-\ihalf V_j}{u_{j,k}+\ihalf V_j}\right)^L\mathop{\prod_{j'=1}^{J}\prod_{k'=1}^{K_{j'}}}_{(j',k')\neq(j,k)}\frac{u_{j,k}-u_{j',k'}+\ihalf M_{j,j'}}{u_{j,k}-u_{j',k'}-\ihalf M_{j,j'}}=1,\qquad
\prod_{j=1}^{J}\prod_{k=1}^{K_j}\frac{u_{j,k}+\ihalf V_j}{u_{j,k}-\ihalf V_j}=1.
\eea
where $J=5$
is the rank of the algebra $osp(6|4)$, $M_{j,j'}$ is the symmetric Cartan matrix and $V_j$ are the Dynkin labels which specify the representation of spin sites. The distinguished simple root system is
\bea\label{DSTS}
\Delta^{0}=\{\de_1-\de_2,\de_2-\ep_1,\ep_1-\ep_2,\ep_2-\ep_3,\ep_2+\ep_3\}.
\eea
we label the simple roots as $\al_1,\al_2,\al_3,\al_4,\al_{\bar4}$ in the given order above. For more details for the algebra $osp(6|4)$, see Appendix~\ref{appendixc}. As shown in Fig.~\ref{dynkin},

\be V_j=(0, 0, 0, 1, 1).  \ee
These equations can be written in a compact form,
\bea \label{BEA}
\mathop{\prod_{j'=0}^{J}\prod_{k'=1}^{K_j'}}_{(j',k')\neq(j,k)}S_{j,j'}(u_{j,k},u_{j',k'})=1,
\eea
with
\bea
S_{j,j'}=\frac{u_{j,k}-u_{j',k'}+\ihalf M_{j,j'}}{u_{j,k}-u_{j',k'}-\ihalf M_{j,j'}},\qquad S_{j,0}=S_{0,j}^{-1}=\frac{u_{j,k}+\ihalf V_j}{u_{j,k}-\ihalf V_j},\qquad S_{0,0}=1, \qquad K_0=L.
\eea

\begin{figure}\label{DDynkin}
\centering
\includegraphics[width=8cm]{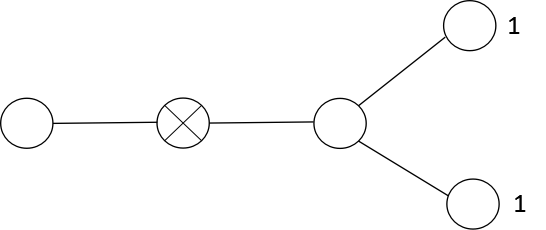}
\caption{The distinguished Dynkin diagram of the algebra $osp(6|4)$.}\label{dynkin}
\end{figure}
\subsection{Orbifolding the Bethe Ansatz}
The leading order orbifold Bethe ansatz equations has the general form \cite{c2} (in the twist $m$ sector for $\mathbb{Z}_n$ orbifold),
\be\label{LOOBAE}
e^{2\pi imq_j/n}\left(\frac{u_{j,k}-\ihalf V_j}{u_{j,k}+\ihalf V_j}\right)^L
\mathop{\prod_{j'=1}^{J}\prod_{k'=1}^{K_j'}}_{(j',k'\neq(j,k))}
\frac{u_{j,k}-u_{j',k'}+\ihalf M_{j,j'}}{u_{j,k}-u_{j',k'}-\ihalf M_{j,j'}}=1,\\
\ee
\be\label{zmc}
e^{2\pi imq_0/n}\prod_{j'=1}^{J}\prod_{k'=1}^{K_{j'}}
\frac{u_{j',k'}+\ihalf V_{j'}}{u_{j',k'}-\ihalf V_{j'}}=1,\\
\ee
\be\label{TwistConstraint}
e^{-2\pi iLq_0/n}\prod_{j'=1}^{J}e^{-2\pi iK_{j'}q_{j'}/n}=1.
\ee
Where the $q_j, j=1,2,3,4,\bar4$ are the $SU(4)$ charges of simple roots under orbifolding, and $q_0$ is the charge of vacuum. In the distinguished simple root system, the charges are,
\be
\bm{q}=\(-t_2-t_3|0,-t_1,2t_1-t_2-t_3,-t_1+2t_2,-t_1+2t_3\).
\ee
with $t_1,t_2,t_3$ integers. The $\bm{q}$ is related the charges $s_I$ by
\be\label{sq}
s_I=\(t_2,t_1-t_2,-t_1+t_3,-t_3\), q_0=s_4-s_1, q_2=-s_1-s_2, q_3=s_2-s_3, q_4=s_1-s_2, q_{\bar4}=s_3-s_4.
\ee
If we restrict to the scalar sector, one can recover the eqs.~(\ref{BAE1}-\ref{BAE3}) from (\ref{LOOBAE}), the eq.~(\ref{TwistConstraint}) is equivalent to the twist constraint (\ref{twistconstraint}) and eq.~(\ref{zmc}) is the zero momentum condition (\ref{ZMC}).

The energy is given by,
\be
E=\lambda^2 \sum_{j=0}^{J}\sum_{k=1}^{K_j}\left(\frac{i}{u_{j,k}+\ihalf V_j}-\frac{i}{u_{j,k}-\ihalf V_j}\right).
\ee
It is useful to represent the twist field $\gamma$ by a new type of quasi-excitation $j=-1$, with excited number $K_{-1}=m$. The phase shift
\be
S_{j,-1}=1/S_{-1,j}=\exp(2\pi iq_j/n),\qquad j=0,\dots J,\qquad
S_{-1,-1}=1.
\ee
Then the leading order Bethe equations for a $\mathbb{Z}_n$ orbifold theory can also be written in a compact form,
\be \label{OBEA}
\mathop{\prod_{j'=-1}^{J}\prod_{k'=1}^{K_j'}}_{(j',k')\neq(j,k)}S_{j,j'}(u_{j,k},u_{j',k'})=1.
\ee
We now consider a simple example in the $SU(2)\times SU(2)$ sector at two loops to verify our orbifold Bethe ansatz. The $SU(2)\times SU(2)$ sector is made of the elementary excitations $(Y^2|Y_3^{\dg})$ on the odd and even sites above the vacuum $\text{Tr}((Y^1Y_4^{\dg})^L)$, and it is closed at any order \cite{GGY, GHO}. At leading order of ABJM theory, the Hamiltonian reduces to the sum of two decoupled Heisenberg $XXX_{1/2}$ Hamiltonians, one acting on the even sites and the other acting on the odd sites\footnote{More precisely speaking, these two chains are only coupled by the zero momentum condition.}
\be
H=\lambda^2\sum_{l=1}^{2L}(1-P_{l,l+2}).
\ee
In orbifold case, the $l$-th term in the Hamiltonian is the same as above for $1\le l\le 2L-2$, and the $2L-1$-th term and the $2L$-th term are multiplied by the phases indicated in eqs.~(\ref{tbc1}) and (\ref{tbc2}), respectively. We consider two excitations above the ``twist vacuum'' $\text{Tr}(\gamma^m(Y^1Y_4^{\dg})^L)$, one on the even sites and another on the odd sites. The obtained operators are  $\text{Tr}(\gamma^m(Y^2Y_3^{\dg})(Y^1Y_4^{\dg})^{L-1})$ and the ones with permutations among even sites and odd sites independently.
For the above operator to be non-vanishing, the twist constraint
\be
\frac{m[(L-1)(s_4-s_1)-s_2+s_3]}{n}\in\mathbb{Z},
\ee
must be imposed.
For concreteness, we take $L=3$. In the basis,
\bea
\ml{O}_1=\text{Tr}(\gamma^mY^2Y_3^{\dg}Y^1Y_4^{\dg}Y^1Y_4^{\dg}), \ml{O}_2=\text{Tr}(\gamma^mY^2Y_4^{\dg}Y^1Y_3^{\dg}Y^1Y_4^{\dg}),
\ml{O}_3=\text{Tr}(\gamma^mY^2Y_4^{\dg}Y^1Y_4^{\dg}Y^1Y_3^{\dg}).
\eea
The Hamiltonian takes the form,
\be
H=\lambda^2\matr{ccc}{
    4&-(1+\omega^{-mq_0})&-\omega^{mq_{\bar4}}(1+\omega^{mq_0})\\
    -(1+\omega^{mq_0})&4&-(1+\omega^{-mq_0})\\
    -\omega^{-mq_{\bar4}}(1+\omega^{-mq_0})&-(1+\omega^{mq_0})&4
  }.
\ee
To write the Hamiltonian in a compact form, we have used the eq.~(\ref{sq}).
With the aid of {\it Mathematica}, it is easy to find the eigenvalues
\be\label{E}
E=4\lambda^2[\text{sin}^2(\frac{m\pi q_{\bar4}}{3n}+\frac{k\pi}{3})+\text{sin}^2(\frac{m\pi (q_{\bar4}+3q_0)}{3n}+\frac{k\pi}{3})], \, k=0,1,2.
\ee
Let's compute it using our orbifold Bethe ansatz equations. In the above simple case $L=3$,  we have all excitation numbers to be zero except $K_4=K_{\bar4}=1$. The Bethe equations are simplified to be
\bea
\(\frac{u+\ihalf}{u-\ihalf}\)^3=\text{e}^{2\pi imq_4/n},\label{u}\\
\(\frac{v+\ihalf}{v-\ihalf}\)^3=\text{e}^{2\pi imq_{\bar4}/n},\label{v}\\
\frac{u+\ihalf}{u-\ihalf}\frac{v+\ihalf}{v-\ihalf}=\text{e}^{-2\pi imq_0/n}.\label{uv}
\eea
However, these three equations are not independent if we impose the twist constraint
\be
\frac{m(3q_0+q_4+q_{\bar4})}{n}\in\mathbb{Z}.
\ee
The energy is given by
\be\label{energy}
E=\lambda^2\(\frac{1}{u^2+\frac14}+\frac{1}{v^2+\frac14}\).
\ee
The solutions of eqs. (\ref{u})-(\ref{uv}) are
\be u=-\frac12\cot\left(\frac{m\pi(3q_0+q_{\bar4})}{3n}+\frac{k\pi}{3}\right), v=\frac12\cot\left(\frac{m\pi q_{\bar4}}{3n}+\frac{k \pi}{3}\right).
\ee
Substituting this in eq.~(\ref{energy}) reproduces the result (\ref{E}) obtained by diagonalising the Hamiltonian directly.
\section{Higher Loops}\label{section5}
We want to generalize our orbifold Bethe equations to higher loops. Firstly, we know the all loop $AdS_4/CFT_3$ asymptotic Bethe equations read \cite{c9},
\be\label{Allloop}
\begin{aligned}
1&=\prod_{j=1}^{K_2}\frac{u_{1,k}-u_{2,j}+\ihalf}{u_{1,k}-u_{2,j}-\ihalf}
\prod_{j=1}^{K_{4}}\frac{1-1/x_{1,k}x^+_{4,j}}{1-1/x_{1,k}x_{4,j}^-}
\prod_{j=1}^{K_{\bar 4}}\frac{1-1/x_{1,k} x^+_{\bar 4,j}}{1-1/x_{1,k}x_{\bar 4,j}^-} \,,\\
1&=\prod_{j=1,j\neq k}^{K_2}\frac{u_{2,k}-u_{2,j}-i }{u_{2,k}-u_{2,j}+i}
\prod_{j=1}^{K_1}\frac{u_{2,k}-u_{1,j}+\ihalf}{u_{2,k}-u_{1,j}-\ihalf}
\prod_{j=1}^{K_3}\frac{u_{2,k}-u_{3,j}+\ihalf}{u_{2,k}-u_{3,j}-\ihalf}\,,\\
1&=\prod_{j=1}^{K_2}\frac{u_{3,k}-u_{2,j}+\ihalf}{u_{3,k}-u_{2,j}-\ihalf}
\prod_{j=1}^{K_4}\frac{x_{3,k}-x^+_{4,j}}{x_{3,k}-x^-_{4,j}}
\prod_{j=1}^{K_{\bar4}}\frac{x_{3,k}-x^+_{\bar4,j}}{x_{3,k}-x^-_{\bar4,j}} \,,\\
\(\frac{x^+_{4,k}}{x^-_{4,k}}\)^{L}
&=\prod_{j=1,j\neq k}^{K_4}\frac{u_{4,k}-u_{4,j}+i}{u_{4,k}-u_{4,j}-i}\,
\prod_{j=1}^{K_1}\frac{1-1/x^-_{4,k}x_{1,j}}{1-1/x^+_{4,k} x_{1,j}}
\prod_{j=1}^{K_3}\frac{x^-_{4,k}-x_{3,j}}{x^+_{4,k}-x_{3,j}}\times\\
&\times\prod_{j=1,j\neq k}^{K_4}\sigma_{BES}(u_{4,k},u_{4,j})
\prod_{j=1}^{K_{\bar4}}\sigma_{BES}(u_{4,k},u_{\bar4,j})\,, \\
\(\frac{x^+_{\bar4,k}}{x^-_{\bar4,k}}\)^{L}
&=\prod_{j=1}^{K_{\bar4}}\frac{u_{\bar4,k}-u_{\bar4,j}+i}{u_{\bar4,k}-u_{\bar4,j}-i}\,
\prod_{j=1}^{K_1}\frac{1-1/x^-_{\bar4,k} x_{1,j}}{1-1/x^+_{\bar4,k} x_{1,j}}
\prod_{j=1}^{K_3}\frac{x^-_{\bar4,k}-x_{3,j} }{x^+_{\bar4,k}-x_{3,j}}\times\\
&\times\prod_{j=1,j\neq k}^{K_{\bar4}}\sigma_{BES}(u_{\bar4,k},u_{\bar4,j})
\prod_{j=1}^{K_{4}}\sigma_{BES}(u_{\bar4,k},u_{4,j})\,.
\end{aligned}
\ee
where the $x^\pm$ are Zhukowski variables,
\be\label{X}
x+\frac{1}{x}=\frac{u}{h(\lambda)}\;\;,\;\;x^\pm+\frac{1}{x^\pm}=\frac{1}{h(\lambda)}\(u\pm\ihalf\) \,.
\ee
Note that the eqs.~(\ref{Allloop}) still have the form of eq.~(\ref{BEA}) except that one uses the rapidities $x_{j,k}$ instead of $u_{j,k}$ and the scattering phases $S_{j,j'}(x_{j,k},x_{j',k'})$ between the various Bethe roots are modified to accommodate the higher-loop interactions.\\

\begin{figure}
  \centering
  \includegraphics[width=8cm]{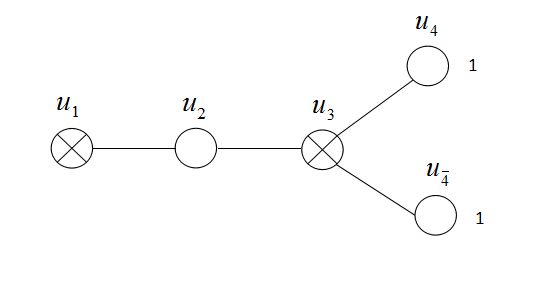}\\
  \caption{The ``higher" Dynkin diagram for $AdS_4/CFT_3$}\label{Dynkin}
\end{figure}

Unlike the leading order Bethe equations, not all simple root systems are possible for writing down higher loops Bethe ansatz equations. One of the possible Dynkin diagrams is shown in Fig.~\ref{Dynkin}.
Another possible ``Higher'' Dynkin diagram is given by performing Weyl reflections with respect to the 1st and the 3rd simple roots in succession, and the result is the diagram on the right side in Fig.~\ref{twograding}. See Appendix~\ref{appendixc} for details. The two corresponding all loop Bethe equations are mapped to each other by ``fermionic duality'' which is consistent with odd Weyl reflection.
The all loop Bethe equations has another  ``dynamic transformation'' symmetry which transform the Bethe roots of type 1 into type 3 and change the spin chain length \cite {c10}.  We will prove these two dualities after we give the all-order Bethe ansatz equations. The study of fermionic duality makes sure that these two simple root systems do give the equivalent BAEs for orbifold ABJM theories and helps us to identify  simple root systems which can be used at all loop level. The valid of dynamic duality admits the dynamical nature of the higher loop BAEs which takes into the fact that some operators with different length can mix with each other at higher loop level.
\subsection{The All loop Orbifold Bethe Equations and Dualites}
\subsubsection{The all loop Orbifold Bethe equations}
The Cartan matrix of two gradings $\eta=\pm1$ can be summarized as
\be
M_{jj'}=\matr{ccccc}{
    &         +\eta&          &         &           \\
  +\eta&     -2\eta&     +\eta&         &           \\
    &         +\eta&          &    -\eta&    -\eta  \\
    &          &         -\eta&   1+\eta&   -1+\eta \\
    &          &         -\eta&  -1+\eta&    1+\eta
  }
\ee
\begin{figure}[H]
  \centering
  \subfigure{
  \begin{minipage}{6cm}
  \includegraphics[width=6cm]{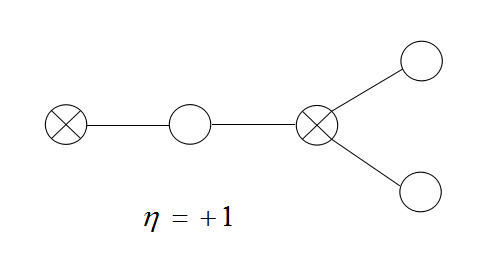}
  \end{minipage}
  }
  \subfigure{
  \begin{minipage}{6cm}
  \includegraphics[width=6cm]{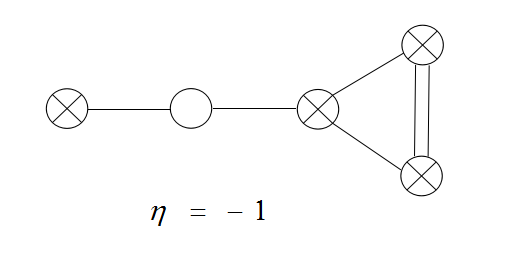}
  \end{minipage}
  }
  \caption{\footnotesize{Two choice of Dynkin daigrams for higher loops Bethe equations.}}
  \label{twograding}
\end{figure}
and the charges for $\eta=+1$,
\be\label{qplus}
\bm{q}^+=(-t_2-t_3|t_1,0,t_1-t_2-t_3,-t_1+2t_2,-t_1+2t_3),
\ee
for $\eta=-1$,
\be\label{qminus}
\bm{q}^-=(-t_2-t_3|-t_1,2t_1-t_2-t_3,-t_1+t_2+t_3,t_2-t_3,-t_2+t_3).
\ee

Now we give a bit details of  the derivation of $\bm{q}^+$, the one of $\bm{q}^-$ is similar.

To do this we need the Cartan matrix for the Dynkin diagram in Fig.~\ref{Dynkin},
\be
M_{jj'}=\matr{ccccc}{
    &         +1 &     &        &         \\
  +1&         -2&    +1 &        &         \\
    &        +1&     &      -1&      -1 \\
    &          &     -1&      +2&         \\
    &          &     -1&        &      +2
  }.\label{cartan2}
\ee
Then we begin with the distinguished simple root system (\ref{DSTS}). First we apply $w_{\alpha_2}$ and give
\be
w_{\alpha_2}(\Delta^0)=\{\delta_1-\ep_1,-\de_2+\ep_1,\de_2-\ep_2,\ep_2-\ep_3,\ep_2+\ep_3\},
\ee
then apply another $w_{\alpha_1}$ with $\alpha_1$ being the first simple root $\de_1-\ep_1$ in new basis
\be
w_{\alpha_1}(w_{\alpha_2}(\Delta^0))=\{-\de_1+\ep_1,\de_1-\de_2,\de_2-\ep_2,\ep_2-\ep_3,\ep_2+\ep_3\}.
\ee
Now we get the ``higher'' simple root system with the Dynkin diagram shown in the Fig.~\ref{Dynkin}.
The original three $SO(6)$ simple roots $\ep_1-\ep_2,\ep_2-\ep_3,\ep_2+\ep_3$ can be found in this basis as
\be
\begin{aligned}
\ep_1-\ep_2=&\alpha_1+\alpha_2+\alpha_3,\\
\ep_2-\ep_3=&\alpha_4,\\
\ep_2+\ep_3=&\alpha_{\bar{4}}.
\end{aligned}
\ee

Now, adding the first and two rows in eq.~(\ref{cartan2}) to the third one, and multiplying the obtained matrix from the right to $(0, 0, t_1, t_2, t_3)$, we get
\be (q_1, q_2, q_3, q_4, q_{\bar{4}})=(t_1,0,t_1-t_2-t_3,-t_1+2t_2,-t_1+2t_3). \ee
and the non-vanishing Dynkin labels are the same with the distinguished simple root system because we have merely dualized the first and the second simple root. Also because of this, $q_0$ does not change and we get eq.(\ref{qplus}).
From eqs.~(\ref{qplus}) and (\ref{qminus}), we observe that
\be\label{ChargeRelation1}
q^{+\eta}_3=q^{+\eta}_1+\eta q^{+\eta}_0.
\ee
\be\label{ChargeRelation2}
q_4^++q_3^+=q_4^-,\, q_{\bar 4}^++q_3^+=q_{\bar 4}^-.
\ee
\be\label{ChargeRelation3}
q_0^{+\eta}-2q_3^+=q_2^+-q_2^-.
\ee
\be\label{ChargeRelation4}
q_1^++q_1^-=0.
\ee
\be\label{ChargeRelation5}
q_3^++q_3^-=0.
\ee
The all loop $AdS_4/CFT_3$  orbifold Bethe equations read,
\be\label{AllloopOrbifold}
\begin{aligned}
e^{-2\pi imq_1^{+\eta}/n}&=\prod_{j=1}^{K_2}\frac{u_{1,k}-u_{2,j}+\ihalf\eta}{u_{1,k}-u_{2,j}-\ihalf\eta}
\prod_{j=1}^{K_4}\frac{1-1/x_{1,k} x^{+\eta}_{4,j}}{1-1/x_{1,k}x_{4,j}^{-\eta}}
\prod_{j=1}^{K_{\bar 4}}\frac{1-1/x_{1,k} x^{+\eta}_{\bar 4,j}}{1-1/x_{1,k}x_{\bar 4,j}^{-\eta}}\,,\\
e^{-2\pi imq_2^{+\eta}/n}&=\prod_{j=1,j\neq k}^{K_2}\frac{u_{2,k}-u_{2,j}-i\eta}{u_{2,k}-u_{2,j}+i\eta}
\prod_{j=1}^{K_1}\frac{u_{2,k}-u_{1,j}+\ihalf\eta}{u_{2,k}-u_{1,j}-\ihalf\eta}
\prod_{j=1}^{K_3}\frac{u_{2,k}-u_{3,j}+\ihalf\eta}{u_{2,k}-u_{3,j}-\ihalf\eta}\,,\\
e^{-2\pi imq_3^{+\eta}/n}&=\prod_{j=1}^{K_2}\frac{u_{3,k}-u_{2,j}+\ihalf\eta}{u_{3,k}-u_{2,j}-\ihalf\eta}
\prod_{j=1}^{K_4}\frac{x_{3,k} -x^{+\eta}_{4,j}}{x_{3,k} -x^{-\eta}_{4,j}}
\prod_{j=1}^{K_{\bar 4}}\frac{x_{3,k} -x^{+\eta}_{\bar 4,j}}{x_{3,k} -x^{-\eta}_{\bar 4,j}} \,,\\
e^{-2\pi imq_4^{+\eta}/n}\(\frac{x^+_{4,k}}{x^-_{4,k}}\)^{L}&=
\prod_{j=1,j\neq k}^{K_4}\frac{x^+_{4,k}-x^-_{4,j}}{x^{-\eta}_{4,k}-x^{+\eta}_{4,j}}
\frac{1-1/x^+_{4,k}x^-_{4,j}}{1-1/x^-_{4,k}x^+_{4,j}}\sigma_{BES}(x_{ 4,k},x_{ 4,j})\,
\prod_{j=1}^{K_1}\frac{1-1/x^{-\eta}_{4,k} x_{1,j}}{1-1/x^{+\eta}_{4,k} x_{1,j}}\\
&\times\prod_{j=1}^{K_3}\frac{x^{-\eta}_{4,k}-x_{3,j} }{x^{+\eta}_{4,k}-x_{3,j}}
\prod_{j=1}^{K_{\bar 4}}\frac{x^{+\eta}_{4,k}-x^+_{\bar4,j}}{x^+_{4,k}-x^{+\eta}_{\bar 4,j}}
\sigma_{BES}(x_{4,k},x_{\bar 4,j})  \,,\\
e^{-2\pi imq_{\bar4}^{+\eta}/n}\(\frac{x^+_{\bar 4,k}}{x^-_{\bar 4,k}}\)^{L}
&=\prod_{j=1,j\neq k}^{K_{\bar 4}}
\frac{x^+_{\bar 4,k}-x^-_{\bar 4,j}}{x^{-\eta}_{\bar 4,k}-x^{+\eta}_{\bar 4,j}}
\frac{1-1/x^+_{\bar 4,k}x^-_{\bar 4,j}}{1-1/x^-_{\bar 4,k}x^+_{\bar 4,j}}
\sigma_{BES}(x_{\bar 4,k},x_{\bar 4,j})\,
\prod_{j=1}^{K_1}\frac{1-1/x^{-\eta}_{\bar 4,k} x_{1,j}}{1-1/x^{+\eta}_{\bar 4,k} x_{1,j}}\\
&\times\prod_{j=1}^{K_3}\frac{x^{-\eta}_{\bar 4,k}-x_{3,j} }{x^{+\eta}_{\bar 4,k}-x_{3,j}}
\prod_{j=1}^{K_{4}}\frac{x^{+\eta}_{\bar4,k}-x^+_{4,j}}{x^+_{\bar4,k}-x^{+\eta}_{4,j}}
\sigma_{BES}(x_{\bar4,k},x_{4,j})  \,.\\
\end{aligned}
\ee
subject to zero momentum constraint,
\be\label{Momentum}
e^{-2\pi imq_0^{+\eta}/n}=\prod_{j=1}^{K_4}\frac{x^+_{4,j}}{x^-_{4,j}}
\prod_{j=1}^{K_{\bar 4}}\frac{x^+_{\bar 4,j}}{x^-_{\bar 4,j}}\,,
\ee
and the twist condition (\ref{TwistConstraint})
 Here $\sigma_{BES}$ is the BES kernel for ABJM theory whose concrete expression can be found in \cite{c9}.
The spectrum of energy is
\be E=\sum_{j=1}^{K_4}\frac12\left(\sqrt{1+16h(\lambda)^2\sin^2\frac{p_j}{2}}-1\right)
+\sum_{j=1}^{K_{\bar{4}}}\frac12\left(\sqrt{1+16h(\lambda)^2\sin^2\frac{\bar{p}_j}{2}}-1\right), \ee
where\be p_j=\frac{1}{i}\log\frac{x_{4, j}^+}{x_{4, j}^-}, \,  \bar{p}_j=\frac{1}{i}\log\frac{x_{\bar{4}, j}^+}{x_{\bar{4}, j}^-}, \ee
and $h(\lambda)$ is an interpolating function \cite{GGY, NT, GHO, Gromov} and it also replaces $\sqrt{\lambda}/(4\pi)$ appearing in the BES kernel for 4d $\ml{N}=4$ Super Yang-Mills theory.
We have the following relations,
\be\label{Xu}
\begin{aligned}
u_k-u_j&=h(\lambda)(x_k-x_j)(1-1/x_kx_j)=h(\lambda)(x^\pm_k-x^\pm_j)(1-1/x^\pm_kx^\pm_j)\,,\\
u_k-u_j\pm\ihalf&=h(\lambda)(x^\pm_k-x_j)(1-1/x^\pm_kx_j)=h(\lambda)(x_k-x^\mp_j)(1-1/x_kx^\mp_j)\,,\\
u_k-u_j\pm i&=h(\lambda)(x^\pm_k-x^\mp_j)(1-1/x^\pm_kx^\mp_j)\,.
\end{aligned}
\ee
They are easily confirmed using the definition (\ref{X}).
\subsubsection{Dynamic Duality}
The equation for $x_3$ is,
\be
e^{-2\pi imq_3^{+\eta}/n}=\prod_{j=1}^{K_2}\frac{u_{3,k}-u_{2,j}+\ihalf\eta}{u_{3,k}-u_{2,j}-\ihalf\eta}
\prod_{j=1}^{K_4}\frac{x_{3,k} -x^{+\eta}_{4,j}}{x_{3,k} -x^{-\eta}_{4,j}}
\prod_{j=1}^{K_{\bar 4}}\frac{x_{3,k} -x^{+\eta}_{\bar 4,j}}{x_{3,k} -x^{-\eta}_{\bar 4,j}} \,.
\ee
We now transform one type 3 root $1/x_{3,k}\rightarrow x_{1,k}$. For this transformation, $u_{3,k}\rightarrow u_{1,k}$, and one of the $x_3$ equations transforms as,
\be
\prod_{j=1}^{K_2}\frac{u_{1,k}-u_{2,j}+\ihalf\eta}{u_{1,k}-u_{2,j}-\ihalf\eta}\prod_{j=1}^{K_4}\frac{x^{+\eta}_{4,j}}{x^{-\eta}_{4,j}}
\frac{1-1/x_{1,k} x^{+\eta}_{4,j}}{1-1/x_{1,k}x_{4,j}^{-\eta}}\prod_{j=1}^{K_{\bar 4}}
\frac{x^{+\eta}_{\bar 4,j}}{x^{-\eta}_{\bar 4,j}}
\frac{1-1/x_{1,k} x^{+\eta}_{\bar 4,j}}{1-1/x_{1,k}x_{\bar 4,j}^{-\eta}}=e^{-2\pi imq_3^{+\eta}/n}\,.
\ee
Using the momentum condition (\ref{Momentum}), we find
\be
\prod_{j=1}^{K_2}\frac{u_{1,k}-u_{2,j}+\ihalf\eta}{u_{1,k}-u_{2,j}-\ihalf\eta}
\prod_{j=1}^{K_4}\frac{1-1/x_{1,k} x^{+\eta}_{4,j}}{1-1/x_{1,k}x_{4,j}^{-\eta}}
\prod_{j=1}^{K_{\bar 4}}\frac{1-1/x_{1,k} x^{+\eta}_{\bar 4,j}}{1-1/x_{1,k}x_{\bar 4,j}^{-\eta}}
=e^{-2\pi imq_3^{+\eta}/n+2\pi im\eta q_0^{+\eta}/n}=e^{-2\pi imq_1^{+\eta}/n}\,.
\ee
where the relation(\ref{ChargeRelation1}) has been used.
We recognize that this is the equation for $x_{1}$ with the same grading.
Under this transformation, the scattering phases in $x_4$ and $x_{\bar 4}$ equations also get changed. For example,
\be
\frac{x^{-\eta}_{4,k}-x_{3,j} }{x^{+\eta}_{4,k}-x_{3,j}}\rightarrow \frac{x^{-\eta}_{4,k}}{x^{+\eta}_{4,k}}
\frac{1-1/x^{-\eta}_{4,k} x_{1,j}}{1-1/x^{+\eta}_{4,k} x_{1,j}}\,,
\ee
and the equation for $x_4$ transforms as,
\be
\begin{aligned}
e^{-2\pi imq_4^{+\eta}/n}\(\frac{x^+_{4,k}}{x^-_{4,k}}\)^{L+\eta}
&=\prod_{j=1,j\neq k}^{K_4}\frac{x^+_{4,k}-x^-_{4,j}}{x^{-\eta}_{4,k}-x^{+\eta}_{4,j}}
\frac{1-1/x^+_{4,k}x^-_{4,j}}{1-1/x^-_{4,k}x^+_{4,j}}
\sigma_{BES}(x_{ 4,k},x_{ 4,j})\,
\prod_{j=1}^{K_1+1}\frac{1-1/x^{-\eta}_{4,k} x_{1,j}}{1-1/x^{+\eta}_{4,k} x_{1,j}}\\
&\times\prod_{j=1}^{K_3-1}\frac{x^{-\eta}_{4,k}-x_{3,j} }{x^{+\eta}_{4,k}-x_{3,j}}
\prod_{j=1}^{K_{\bar 4}}\frac{x^{+\eta}_{4,k}-x_{ \bar 4,j}}{x^+_{4,k}-x^{+\eta}_{\bar 4,j}}
\sigma_{BES}(x_{4,k},x_{\bar 4,j})  \,,\\
\end{aligned}
\ee
thus with this transformation in addition to the following replacements which is called dynamic transformation,
\be
K_3\rightarrow K_3-1, K_1\rightarrow K_1+1, L\rightarrow L+\eta\,.
\ee
the all loop Bethe eqs.~(\ref{Allloop}) remain invariant with the same grading. The momentum conservation condition and the expression for the total energy are not changed under the dynamic duality.

This dynamic duality is closely related to properties of the all-loop S-matrix. We only demonstrate this for grading $\eta=1$.
In fact, for the ABJM case, the needed property is
\be \label{Phase}
S_{j,3}(x,x_3)=S_{j,1}(x,x_1)S_{j,0}(x)\qquad\text{for}\;j=1,2,3,4,\bar{4}.
\ee
for $x_1x_3=1$.
This can be checked directly. We only give the  proof for  the case with $j=4,\bar{4}$, other cases are trivial.
\be
\begin{aligned}
S_{4,3}=\frac{x_4^--x_3}{x_4^+-x_3},\qquad S_{4,1}=\frac{1-1/x_4^-x_1}{1-1/x_4^+x_1},\qquad S_{4,0}=\frac{x_4^-}{x_4^+}.\\
\end{aligned}
\ee
Using $x_3x_1=1$, we find,
\be
\frac{x_4^--1/x_1}{x_4^+-1/x_1}=\frac{x_4^-}{x_4^+}\frac{1-1/x_4^-x_1}{1-1/x_4^+x_1}\,.
\ee
The case for $j=\bar{4}$ is similar.
For the orbifold theories, we need the relation(\ref{Phase}) holds for $j=-1$ as well. This is the case because $q_3^+=t_1-t_2-t_3=q_1^++q_0^+$.

\subsubsection{The Fermionic Duality}
We now prove that two choices of grading in (\ref{AllloopOrbifold}) are equivalent based on some fermionic duality. In order to investigate this duality of the eqs.(\ref{AllloopOrbifold}), we rewrite the equation of $x_3$ for $\eta=+1$ as,
\be\label{X3}
\prod_{j=1}^{K_2}\frac{x_{3,k}-x^-_{2,j}}{x_{3,k}-x^+_{2,j}}
\prod_{j=1}^{K_2}\frac{x_{3,k}-1/x^-_{2,j}}{x_{3,k}-1/x^+_{2,j}}
\prod_{j=1}^{K_4}\frac{x_{3,k} -x^{+}_{4,j}}{x_{3,k} -x^{-}_{4,j}}
\prod_{j=1}^{K_{\bar 4}}\frac{x_{3,k} -x^{+}_{\bar 4,j}}{x_{3,k} -x^{-}_{\bar 4,j}}
=e^{-2\pi imq_3^{+\eta}/n}\,.
\ee
We further introduce the following polynomial $P(x)$,
\be\label{P}
\begin{aligned}
P(x)&=e^{2\pi imq_3^{+}/n}\prod_{j=1}^{K_2}(x-x^-_{2,j})
\prod_{j=1}^{K_2}(x-1/x^-_{2,j})
\prod_{j=1}^{K_4}(x -x^{+}_{4,j})
\prod_{j=1}^{K_{\bar 4}}(x-x^{+}_{\bar 4,j})\\
&-\prod_{j=1}^{K_2}(x-x^+_{2,j})
\prod_{j=1}^{K_2}(x-1/x^+_{2,j})
\prod_{j=1}^{K_4}(x -x^{-}_{4,j})
\prod_{j=1}^{K_{\bar 4}}(x-x^{-}_{\bar 4,j})\,.
\end{aligned}
\ee
Obviously there already exists $K_3+K_1$ roots of $P(x)$,
\be
P(x_{3,k})=0,k=1,\dots, K_3,\qquad P(1/x_{1,k})=0, k=1, \dots, K_1\,.
\ee
The remaining solutions can also be grouped into two classes, type 3 roots and type 1 roots.
\be\label{P1}
P(x)\sim \prod_{j=1}^{K_3}(x-x_{3,j})\prod_{j=1}^{K_1}(x-1/x_{1,j})\prod_{j=1}^{\tilde{K}_3}(x-\tilde{x}_{3,j})\prod_{j=1}^{\tilde{K}_1}(x-1/\tilde{x}_{1,j})\,,
\ee
where
\be\label{Tilde}
\tilde{K}_3=K_2+K_4+K_{\bar4}-K_3,\qquad \tilde{K}_1=K_2-K_1\,.
\ee
We now calculate $P(x^\pm_{4,k})$ using two equivalent expressions of $P(x)$,
\be
\begin{aligned}
\frac{P(x^+_{4,k})}{P(x^-_{4,k})}&=
e^{-2\pi imq_3^{+}/n}
\prod_{j=1,j\neq k}^{K_4}\frac{x^+_{4,k}-x^-_{4,j}}{x^-_{4,k}-x^+_{4,j}}
\prod_{j=1 }^{K_{\bar4}}\frac{x^+_{4,k}-x^-_{\bar4,j}}{x^-_{4,k}-x^+_{\bar4,j}}
\prod_{j=1}^{K_2}\frac{x^+_{4,k}-x^+_{2,j}}{x^-_{4,k}-x^-_{2,j}}
\prod_{j=1}^{K_2}\frac{x^+_{4,k}-1/x^+_{2,j}}{x^-_{4,k}-1/x^-_{2,j}}\\
&=\prod_{j=1}^{K_3}\frac{x^+_{4,k}-x_{3,j}}{x^-_{4,k}-x_{3,j}}
\prod_{j=1}^{K_1}\frac{x^+_{4,k}-1/x_{1,j}}{x^-_{4,k}-1/x_{1,j}}
\prod_{j=1}^{\tilde{K}_3}\frac{x^+_{4,k}-\tilde{x}_{3,j}}{x^-_{4,k}-\tilde{x}_{3,j}}
\prod_{j=1}^{\tilde{K}_1}\frac{x^+_{4,k}-1/\tilde{x}_{1,j}}{x^-_{4,k}-1/\tilde{x}_{1,j}}\,.
\end{aligned}
\ee
Using the relations (\ref{Xu}) and (\ref{ChargeRelation2}), we find,
\be
\begin{aligned}
&e^{-2\pi imq_3^{+}/n}\prod_{j=1,j\neq k}^{K_4}\frac{x^+_{4,k}-x^-_{4,j}}{x^-_{4,k}-x^+_{4,j}}
\prod_{j=1}^{K_{\bar4}}\frac{x^+_{4,k}-x^-_{\bar4,j}}{x^-_{4,k}-x^+_{\bar4,j}}
\prod_{j=1}^{K_2}\frac{x^+_{4,j}}{x^-_{4,j}}
=\prod_{j=1}^{K_1+\tilde{K}_1}\frac{x^+_{4,j}}{x^-_{4,j}}
\prod_{j=1}^{K_3}\frac{x^+_{4,k}-x_{3,j}}{x^-_{4,k}-x_{3,j}}\\
&\times \prod_{j=1}^{\tilde{K}_3}\frac{x^+_{4,k}-\tilde{x}_{3,j}}{x^-_{4,k}-\tilde{x}_{3,j}}
\prod_{j=1}^{K_1}\frac{1-1/x_{1,j}x^+_{4,k}}{1-1/x_{1,j}x^-_{4,k}}
\prod_{j=1}^{\tilde{K}_1}\frac{1-1/\tilde{x}_{1,j}x^+_{4,k}}{1-1/\tilde{x}_{1,j}x^-_{4,k}}\,.
\end{aligned}
\ee
Using the relation~(\ref{Tilde}), we arrive at,
\be
\begin{aligned}
&e^{-2\pi imq_3^{+}/n}\prod_{j=1,j\neq k}^{K_4}\frac{x^+_{4,k}-x^-_{4,j}}{x^-_{4,k}-x^+_{4,j}}
\prod_{j=1}^{K_3}\frac{x^-_{4,k}-x_{3,j}}{x^+_{4,k}-x_{3,j}}
\prod_{j=1}^{K_1}\frac{1-1/x_{1,j}x^-_{4,k}}{1-1/x_{1,j}x^+_{4,k}}
=\prod_{j=1}^{\tilde{K}_3}\frac{x^+_{4,k}-\tilde{x}_{3,j}}{x^-_{4,k}-\tilde{x}_{3,j}}\\
&\times\prod_{j=1}^{\tilde{K}_1}\frac{1-1/\tilde{x}_{1,j}x^+_{4,k}}{1-1/\tilde{x}_{1,j}x^-_{4,k}}
\prod_{j=1 }^{K_{\bar4}}\frac{x^-_{4,k}-x^+_{\bar4,j}}{x^+_{4,k}-x^-_{\bar4,j}}\,.
\end{aligned}
\ee
Thus the equation for $x_4$ in the grading $\eta=+1$ is equivalent to one in the grading $\eta=-1$  and similar calculations can be done to show the equivalence of two gradings for $x_{\bar 4}$ equation.
It still remains to prove the equivalence for other equations. For this purpose, we calculate the combination $P(x^{\pm}_{2,k})P(1/x^{\pm}_{2,k})$in two ways,
\be
\begin{aligned}
&\frac{P(x^+_{2,k})}{P(x^-_{2,k})}\frac{P(1/x^+_{2,k})}{P(1/x^-_{2,k})}
=e^{4\pi imq_3^{+}/n}\prod_{j=1,j\neq k}^{K_2}\frac{x^+_{2,k}-x^-_{2,j}}{x^-_{2,k}-x^+_{2,j}}
\prod_{j=1,j\neq k}^{K_2}\frac{x^+_{2,k}-1/x^-_{2,j}}{x^-_{2,k}-1/x^+_{2,j}}
e^{-2\pi imq_0^{+\eta}/n}\\
&\times\prod_{j=1}^{K_2}\frac{1/x^+_{2,k}-x^-_{2,j}}{1/x^-_{2,k}-x^+_{2,j}}
\prod_{j=1,j\neq k}^{K_2}\frac{1/x^+_{2,k}-1/x^-_{2,j}}{1/x^-_{2,k}-1/x^+_{2,j}}
=e^{-2\pi im(q_0^{+\eta}-2q_3^+)/n}\(\prod_{j=1,j\neq k}^{K_2}\frac{u_{2,k}-u_{2,j}+i}{u_{2,k}-u_{2,j}-i}\)^2\\
&=\prod_{j=1}^{K_3}\frac{x^+_{2,k}-x_{3,j}}{x^-_{2,k}-x_{3,j}}\frac{1/x^+_{2,k}-x_{3,j}}{1/x^-_{2,k}-x_{3,j}}
\prod_{j=1}^{K_1}\frac{x^+_{2,k}-1/x_{1,j}}{x^-_{2,k}-1/x_{1,j}}\frac{1/x^+_{2,k}-1/x_{1,j}}{1/x^-_{2,k}-1/x_{1,j}}\\
&\times\prod_{j=1}^{\tilde{K}_3}\frac{x^+_{2,k}-\tilde{x}_{3,j}}{x^-_{2,k}-\tilde{x}_{3,j}}\frac{1/x^+_{2,k}-\tilde{x}_{3,j}}{1/x^-_{2,k}-\tilde{x}_{3,j}}
\prod_{j=1}^{\tilde{K}_1}\frac{x^+_{2,k}-1/\tilde{x}_{1,j}}{x^-_{2,k}-1/\tilde{x}_{1,j}}\frac{1/x^+_{2,k}-1/\tilde{x}_{1,j}}{1/x^-_{2,k}-1/\tilde{x}_{1,j}}\,.\\
\end{aligned}
\ee
Using the relations (\ref{Xu}) and(\ref{ChargeRelation3}), we can rewrite the above equation as,
\be
\begin{aligned}
&e^{2\pi imq_2^{+}/n}\prod_{j=1,j\neq k}^{K_2}\frac{u_{2,k}-u_{2,j}-i}{u_{2,k}-u_{2,j}+i}
\prod_{j=1}^{K_1}\frac{u_{2,k}-u_{1,j}+\ihalf}{u_{2,k}-u_{1,j}-\ihalf}
\prod_{j=1}^{K_3}\frac{u_{2,k}-u_{3,j}+\ihalf}{u_{2,k}-u_{3,j}-\ihalf}\\
&=e^{2\pi imq_2^{-}/n}\prod_{j=1,j\neq k}^{K_2}\frac{u_{2,k}-u_{2,j}+i}{u_{2,k}-u_{2,j}-i}
\prod_{j=1}^{\tilde{K}_1}\frac{u_{2,k}-\tilde{u}_{1,j}-\ihalf}{u_{2,k}-\tilde{u}_{1,j}+\ihalf}
\prod_{j=1}^{\tilde{K}_3}\frac{u_{2,k}-\tilde{u}_{3,j}-\ihalf}{u_{2,k}-\tilde{u}_{3,j}+\ihalf}\,.\\
\end{aligned}
\ee
which proves the equivalence of two gradings of the type 2 equation. From eq.(\ref{P1}), we know that $P(\tilde{x}_{3,k})=P(1/\tilde{x}_{1,k})=0$. While substituting back to (\ref{P}), we find $\tilde{x}_3,\tilde{x}_1$ satisfy the same equation as $x_3,x_1$. Now flip the fractions in the $\tilde{x}_3$ and $\tilde{x}_1$ equations, while using the relations~(\ref{ChargeRelation4}-\ref{ChargeRelation5}), we get equations for the alternative grading. In the end, we have proved that (\ref{Allloop}) are equivalent for the two choices of grading.  Because $q_0^+=q_0^-$,  the momentum conservation condition and the expression for the total energy are not changed under the fermionic duality as well. Notice that the relations among charges (\ref{ChargeRelation1}-\ref{ChargeRelation5}) play important roles in the verification of fermionic duality and dynamics duality. These relations  are automatically satisfied by the charges calculated from the Cartan matrices, instead of imposing by hands in the twisted Bethe ansatz equations studied in \cite{Gromov:2007ky}. In this sense, the check of these two duality for orbifold ABJM theories is a non-trivial check of these all loop BAEs, especially the computations of these charges.

\subsection{Two applications}
As an application, we now compare our results with the all-loop BAE equations for the $\beta$-deformed ABJM theory \cite{chenliuwu}. For $\eta=1$ grading, the phase factors appearing in eqs.~(\ref{Momentum}) and (\ref{AllloopOrbifold})
\bea
-2\pi i \bm{q}^+m/n&=&(2\pi i(t_2+t_3)m/n|-2\pi i t_1m/n, 0, -2\pi i (t_1-t_2-t_3)m/n,\nonumber\\
 &&-2\pi i (-t_1+2 t_2)m/n, -2\pi i (-t_1+2 t_3)m/n), \eea
are replaced by
\bea (-\pi i \beta (K_4-K_{\bar{4}})|0, 0, -\pi i \beta (K_4-K_{\bar{4}}), \pi i \beta(K_3-2 K_{\bar{4}}+L), -\pi i \beta (K_3-2 K_4+L)). \eea
It is easy to see that if \bea t_1&=&0,\\t_2&=&-\frac{n}{4m}\beta (K_3-2K_{\bar{4}}+L), \\ t_3&=&\frac{n}{4m}\beta (K_3-2 K_4+L), \eea
these two groups of phases are the same.  This means if these conditions are satisfied, the all-loop BAEs (for $\eta=1$ grading) for orbifold ABJM theories and $\beta$-deformed ABJM theory coincide for states with these excitation numbers. Notice here $\beta$ should be a rational number and $t_1$ should vanish.
As will be discussed in the next section, $t_1=0$ is the condition for the orbifold  theory to have at least $\ml{N}=2$ supersymmetry. This condition is not surprising since the $\beta$-deformed ABJM theory is $\ml{N}=2$ supersymmetric \cite{Imeroni:2008cr}.

We now turn to relation between cusp anomalous dimension in orbifold ABJM theories and orbifold ${\ml{N}}=4$ SYM theories. As in \cite{c9}, we start with grading $\eta=-1$ and focus on the solutions to all loop BAEs  with only non-vanishing roots $u_{4, k}=u_{\bar{4}, k}$. Then the consistency of the BAEs leads to $q_4^-=q_{\bar{4}}^-$. However from eq.~(\ref{qminus}), we have already $q_4^-=t_2-t_3=-q_{\bar{4}}^-$. Then we are restricted to the case with $q_4^-=q_{\bar{4}}^-=0$. We also demand the phase in the zero momentum condition is trivial,
\be \exp(2\pi i m q_0^-/n)=1.\ee
The above conditions leads to \be t_2=t_3, \, \exp(4\pi i mt_2/n)=1. \label{condition1} \ee
As for the orbifold SYM side with $sl(2)$ grading (corresponding to $\eta_1=\eta_2=-1$ in \cite{Beisert:2005fw}), the phase for the momentum-carrying node is automatically zero (see eq.~(3.15) of \cite{Beccaria:2011qd}).  The triviality of the phase in the zero momentum condition gives
\be \exp(2\pi imt_2^{SYM}/n)=1, \label{condition2}\ee
where $t_2^{SYM}$ is one of the parameters appearing in the orbifold SYM theory.
Under the conditions in eqs.~(\ref{condition1}-\ref{condition2}), we can get the following relations
\be f_{\text{orb. ABJM}}(\lambda)=\frac12f_{\text{orb. SYM}}(\lambda)|_{\frac{\sqrt{\lambda}}{4\pi}\to h(\lambda)}, \ee
as the one obtained in \cite{c9}, under the assumption that wrapping contributions for twist operators are still subleading in the large spin limit with twist being finite.

\section{Supersymmetric orbifold theories}
Let us finally discuss the supersymmetric orbifold theories. Based on the results in previous sections, all we need is to determine the $t_i$'s (or equivalently $\bm{q}$'s in the distinguished simple root system) which are compatible with certain number of supersymmetries. Here we follow the argument of \cite{c2}. We also check the result by determining the spinors of $SO(8)$ preserved by the orbifolding.

\subsection{$\ml{N}=2$ Orbifolds}
To get an $\ml{N}=2$ theory, we need at least one fermionic (odd) generator commuting with the orbifold action. For example, when considering $E^{\al_2}$ corresponding to the only odd simple root $\al_2=-\ep_1+\de_2$ in the distinguished simple root system, this is equivalent to set $q_2=-t_1=0$.
We now demonstrate that this is enough. $\Gamma$ can be naturally embedded into a $U(1)$ subgroup of $OSp(6|4)$. Denote the generator of this $U(1)$ as $\ml{P}$, we have that $\ml{P}^{\dagger}=\ml{P}$. $q_2=0$ means $[E^{-\ep_1+\de_2},\ml{P}]=0$. Then $[E^{\ep_1-\de_2},\ml{P}]=0$, as $(E^{-\ep_1+\de_2})^{\dagger}=E^{\ep_1-\de_2}$.
Note that $E^{\ep_1-\de_2}$ locates in the first line of weight diagram in Fig.~\ref{Odd}, and $E^{-\ep_1+\de_2}$ locates in the last line. By using $Sp(4)$ invariance\footnote{Precisely speaking, this $Sp(4)$ is in fact $Sp(2, 2)$ which is the double cover of $SO(2, 3)$, conformal group of three dimensional spacetime.}, we obtain that all generators in these two lines commute with $\ml{P}$, then we get an $\ml{N}=2$ theory with the charges,
\be
\bm{q}=(-t_2-t_3|0, 0; -t_2-t_3, 2t_2, 2t_3).
\ee
We can also get the charges of $\bm{4}$ of $SO(6)$ as $(t_2,-t_2,t_3,-t_3)$, where $t_2, t_3$ are arbitrary integers except ones satisfying $t_2\pm t_3=0$, because such $t_1, t_2$ will give ${\cal N}=4$ supersymmetries, as we will show below.
This includes the chiral orbifold theory in \cite{c1} as a special case.
We further demonstrate this method is indeed correct by counting the spinors of $SO(8)$ preserved by the orbifold.
Notice that the orbifold ABJM theory is the low energy effective theory of $N$ coincident M2-branes at  $\mathbb{C}^4/(\Gamma\times \mathbb{Z}_{|\Gamma|k})$ orbifold singularity where $|\Gamma|$ is the order of $\Gamma$
and $\mathbb{Z}_{|\Gamma|k}$ acts as overall phase rotations of the four complex coordinates \cite{Berenstein}.
Under the action of the generator of finite group $\Gamma$ as $(Y^1,Y^2,Y^3,Y^4)\rightarrow(\omega^{t_2}Y^1,\omega^{-t_2}Y^2,\omega^{t_3}Y^3,\omega^{-t_3}Y^4)$,
the $SO(8)$ spinor transforms like $\ep\rightarrow\omega^{(\mathfrak{s_1}t_2-\mathfrak{s_2}t_2+\mathfrak{s_3}t_3-\mathfrak{s_4}t_3)}\ep$, where $\mathfrak{s}_{1,2,3,4}=\pm 1/2$.
The equation
\be
\mathfrak{s_1}t_2-\mathfrak{s_2}t_2+\mathfrak{s_3}t_3-\mathfrak{s_4}t_3\in n\mathbb{Z}
\ee
subject to
\be
\mathfrak{s_1}+\mathfrak{s_2}+\mathfrak{s_3}+\mathfrak{s_4} \in k|\Gamma|\mathbb{Z}
\ee
has exact two solutions
\be (\mathfrak{s_1}, \mathfrak{s_2}, \mathfrak{s_3}, \mathfrak{s_4})=\pm(1/2, 1/2, -1/2, -1/2), \ee
for generic $t_2, t_3$ and $n$, and this demonstrates our conclusion above\footnote{We assume $k\geq 3$ here.}.
\subsection{$\ml{N}=4$ Orbifold}
From the above example, we note supersymmetric orbifold ABJM theories always preserve an $\mathcal{N}=even$ supersymmetry as an consequence of the special structure of $osp(6|4)$ algebra while which is not the case in orbifolds of $\ml{N}=4$ SYM theory. This results can be confirmed by spinor counting. If $(\mathfrak{s_1}, \mathfrak{s_2}, \mathfrak{s_3}, \mathfrak{s_4})$ satisfies the projection condition, so does $(-\mathfrak{s_1}, -\mathfrak{s_2}, -\mathfrak{s_3}, -\mathfrak{s_4})$.
We now consider $\ml{N}=4$ orbifold. To find the conditions, without loss of generality we can first demand $q_2=-t_1=0$, then further demand $q_3=2t_1-t_2-t_3=0$, while $q_4=-t_1+2t_2\neq0$,\,$q_{\bar4}=-t_1+2t_3\neq0$. We then have $[E^{\al_3},\ml{P}]=0$, then $[\[E^{-\ep_1+\de_2},E^{\al_3}\],\ml{P}]=0$, this is $[E^{-\ep_2+\de_2},\ml{P}]=0$ and according to the argument in $\ml{N}=2$ case, we further have $[E^{\ep_1-\de_2},\ml{P}]=0$ and $[E^{\ep_2-\de_2},\ml{P}]=0$, together with the $Sp(4)$ symmetry we get an $\ml{N}=4$ theory. Solving these constraints including $q_2=-t_1=0$, we find
$\bm{q}=(0|0, 0; 0, 2t_2, -2t_2)$, with the charges of $\bm{4}$ as $(t_2, -t_2, -t_2, t_2)$.
Using the spinor counting method we can also demonstrate our conclusion is correct.


\section{Discussions}
In this paper, we studied the integrability of planar orbifold ABJM theories. We first carried out perturbative computations of ADM
in the scalar sector at two-loop order. We found that in the corresponding spin chain Hamiltonian, only two terms are deformed by certain phases. This deformation can be expressed in terms of twisted boundary condition. By inserting certain diagonal matrices inside the transfer matrices, we proved the integrability of this Hamiltonian. BAEs and eigenvalues of ADM were obtained through algebraic Bethe ansatz method. Restricting $\Gamma$ to be inside $SU(4)_R$, we obtained the all-loop all-sector BAEs which pass some non-trivial consistency checks. 

There are several interesting directions worth pursuing. One of them is that to explore all-loop BAEs for general $\Gamma$ in $SU(4)_R\times U(1)_b$. This study is beyond the framework of Beisert-Roiban \cite{c2} since $U(1)_b$ does not correspond to a node in the Dynkin diagram used for BAEs. To obtain some hints for the structure of the result, it may be helpful to first perturbatively compute the ADM of composite operators involving fermions as the computation in ABJM theory \cite{Minahan:2009te}. It is also interesting to find some solutions in the thermodynamical limit and study their holographic dual in term of semi-classical string/membrane solutions in the dual string/M theories.

Supersymmetric condition for the orbifold was studied in this framework of integrability. The obtained condition is consistent with the result that orbifold ABJM theory is the low energy effective theory of $N$ membranes put at $\mathbb{C}^4/(\Gamma\times \mathbb{Z}_{|\Gamma|k})$ \cite{Berenstein}. However the study in the integrability side seems only give condition for $\mathbb{Z}_n$ orbifolds which is  $\ml{N}=2$ or $\ml{N}=4$ simultaneously for all $n$. Let us consider the following examples taken from \cite{Terashima:2008ba}. Take $n$ to be even. The cases with  $(t_1, t_2, t_3, t_4)=\pm(n/2, n/2, (-1)^l n, 0), l=0, 1$ is $\ml{N}=2$ supersymmetric and the case with $(t_1, t_2, t_3, t_4)=\pm(n/2, -n/2, (-1)^l n, 0), l=0, 1$ is $\ml{N}=4$ supersymmetric. The preserved supersymmetries can be easily obtained by counting the $SO(8)$ spinors
preserved by the orbifolds. Also notice that all these cases satisfy $\Gamma<SU(4)_R$. We speculate  that these cases do not appear in the analysis of supersymmetric orbifold here because they only appear for even $n$,
not for all integer $n$. It is still interesting to see whether we can probe such cases through some refinements of the studies here. We leave this and directions mentioned previously as suggestions for further studies.

\section*{Acknowledgments}
It is our great pleasure to thank Bin Chen, Jun-Peng Cao, Jian-Xin Lu, Wei Song, Yu-Peng Wang, Gang Yang, Wen-Li Yang, Konstantinos Zoubos for very helpful discussions. We would like to express our special thanks to the anonymous referees for valuable suggestions to improve the paper.
JW would also like to thank the participants of the advanced workshop ``Dark Energy and Fundamental Theory'' supported by the Special Fund for Theoretical Physics from NSFC with Grant No.~11447613 for stimulating discussion. We  thank Institute of Modern Physics, Northwest University and School of Physics and Astronomy, Sun Yat-Sen University for  hospitality in visits during this project. This work was in part supported by NSFC Grant  No. 11575202(NB, HHC, DSL, JW), No. 11475116(XCD) and No. 11222549 (NB, HHC, DSL, JW). JW also gratefully acknowledges the support of K.~C.~Wong Education Foundation.

\begin{appendix}
\section{Hamiltonian of the twisted spin chain}\label{appendixa}
In this appendix, we give the detailed derivation of eq.~(\ref{totalH}). We employ a new set of indices $i=1,2,\cdots ,2L$ to relabel the quantum spaces of the alternating spin chain. Then the monodromy matrices in eqs.~(\ref{mono1}) and (\ref{mono2}) are rewritten as
\bea
T_0(u)=M_0 R_{01}(u)R_{02}(u-2)R_{03}(u)R_{04}(u-2)\cdots R_{0,2L-1}(u)R_{0,2L}(u-2), \\
T_{\bar{0}}(u)=\bar{M}_{\bar{0}} R_{{\bar{0}}1}(u-2)R_{{\bar{0}}2}(u)R_{{\bar{0}}3}(u-2)R_{{\bar{0}}4}(u)\cdots R_{{\bar{0}},2L-1}(u-2)R_{{\bar{0}},2L}(u).
\eea
At the special point $u=0$, the transfer matrices become
\bea
{\tau}(0)&=&\tr_0 (-)^L M_0 P_{01}(-2+K_{02})\cdots P_{0,2L-1}(-2+K_{0,2L})\\\no
&=&\tr_0 (-)^L M_0 (-2+K_{12}) P_{13}(-2+K_{14})\cdots P_{1,2L-1}(-2+K_{1,2L})P_{01}\\\no
&=&(-)^L(-2+K_{12})\prod_{j=2}^LP_{1,2j-1}(-2+K_{1,2j})M_1,\\
\bar{\tau}(0)&=&\tr_{\bar{0}} (-)^L \bar{M}_{\bar{0}}(-2+K_{\bar{0}1})P_{\bar{0}2}\cdots(-2+K_{\bar{0},2L-1})P_{\bar{0},2L}\\\no
&=&\tr_{\bar{0}} (-)^L \bar{M}_{\bar{0}}P_{\bar{0},2L}(-2+K_{2L,1})P_{2L,2}\cdots (-2+K_{2L,2L-1})\\\no
&=&(-)^L\bar{M}_{2L}\prod_{j=1}^{L-1}(-2+K_{2L,2j-1})P_{2L,2j}(-2+K_{2L,2L-1}).
\eea
and
\bea \label{expan1}
&&\frac{d}{du}\tau(u)|_{u=0}\\\no
&=&\tr_0 \sum_{i=1}^{L-1}M_0\left(\prod_{j=1}^{i-1}(-P_{0,2j-1})(-2+K_{0,2j})\right)(-2+K_{0,2i})\left(\prod_{k=i+1}^{L}(-P_{0,2k-1})(-2+K_{0,2k})\right)\\\no
&+&\tr_0\sum_{i=1}^{L-1}M_0\left(\prod_{j=1}^{i-1}(-P_{0,2j-1})(-2+K_{0,2j})\right)(-P_{0,2i-1})\left(\prod_{k=i+1}^{L}(-P_{0,2k-1})(-2+K_{0,2k})\right)\\\no
&+&\tr_0M_0\left(\prod_{j=1}^{L-1}(-P_{0,2j-1})(-2+K_{0,2j})\right)(-2+K_{0,2L})\\\no
&+&\tr_0M_0\left(\prod_{j=1}^{L-1}(-P_{0,2j-1})(-2+K_{0,2j})\right)(-P_{0,2L-1})\\\no
&=&\Sigma_1+\Sigma_2+\Sigma_3+\Sigma_4,
\eea
\bea  \label{expan2}
&&\frac{d}{du}\bar{\tau}(u)|_{u=0}\\\no
&=&\tr_{\bar{0}} \sum_{i=1}^{L-1}\bar{M}_{\bar{0}}\left(\prod_{j=1}^{i-1}(-2+K_{\bar{0},2j-1})(-P_{\bar{0},2j})\right)(-2+K_{\bar{0},2i-1})\left(\prod_{k=i+1}^{L}(-2+K_{\bar{0},2k-1})(-P_{\bar{0},2k})\right)\\\no
&+&\tr_{\bar{0}} \sum_{i=1}^{L-1}\bar{M}_{\bar{0}}\left(\prod_{j=1}^{i-1}(-2+K_{\bar{0},2j-1})(-P_{\bar{0},2j})\right)(-P_{\bar{0},2i})\left(\prod_{k=i+1}^{L}(-2+K_{\bar{0},2k-1})(-P_{\bar{0},2k})\right)\\\no
&+&\tr_{\bar{0}} \bar{M}_{\bar{0}}\left(\prod_{j=1}^{L-1}(-2+K_{\bar{0},2j-1})(-P_{\bar{0},2j})\right)(-2+K_{\bar{0},2L-1})\\\no
&+&\tr_{\bar{0}} \bar{M}_{\bar{0}}\left(\prod_{j=1}^{L-1}(-2+K_{\bar{0},2j-1})(-P_{\bar{0},2j})\right)(-P_{\bar{0},2L})\\\no
&=&\bar{\Sigma}_1+\bar{\Sigma}_2+\bar{\Sigma}_3+\bar{\Sigma}_4,
\eea
where we use $\Sigma_i$,$\bar{\Sigma}_i$,i=1,$\cdots 4$ to label each part in eq.(\ref{expan1}) and (\ref{expan2}) for the convenience of writing. Then let us first deal with $\tau'(0)$ and give some intermediate results of the calculations
\bea\no
\Sigma_1&=&(-)^{L-1}\sum_{i=1}^{L-1}(-2+K_{12})\prod_{j=2}^{i-1}P_{1,2j-1}(-2+K_{1,2j})(-2+K_{1,2i})\prod_{k=i+1}^{L}P_{1,2k-1}(-2+K_{1,2k})M_1,\\
\Sigma_2&=&(-)^{L}\sum_{i=1}^{L-1}(-2+K_{12})\prod_{j=2}^{i-1}P_{1,2j-1}(-2+K_{1,2j})P_{1,2i-1}\prod_{k=i+1}^{L}P_{1,2k-1}(-2+K_{1,2k})M_1.
\eea
\bea
\Sigma_3&=&(-)^{L-1}(-2+K_{12})\prod_{j=2}^{L-1}P_{1,2j-1}(-2+K_{1,2j})(-2+K_{1,2L})M_1,\\
\Sigma_4&=&(-)^{L}(-2+K_{12})\prod_{j=2}^{L-1}P_{1,2j-1}(-2+K_{1,2j})P_{1,2L-1}M_1.
\eea
Thus we find
\bea
\tau(0)^{-1}\Sigma_1&=&-\sum_{i=1}^{L-1}\left(P_{2i-1,2i+1}-\frac{1}{2}K_{2i-1,2i}K_{2i,2i+1}-\frac{1}{2}K_{2i,2i+1}K_{2i-1,2i}+\frac{1}{4}K_{2i,2i+1}\right), \label{ad1}\\
\tau(0)^{-1}\Sigma_2&=&\sum_{i=1}^{L-1}\left(-\frac{1}{2}+\frac{1}{4}K_{2i,2i+1}\right),\\
\tau(0)^{-1}\Sigma_3&=&-M^{-1}_1\left(P_{1,2L-1}-\frac{1}{2}K_{2L-1,2L}K_{1,2L}-\frac{1}{2}K_{1,2L}K_{2L-1,2L}+\frac{1}{4}K_{1,2L}\right)M_1,\\
\tau(0)^{-1}\Sigma_4&=&M_1^{-1}\left(-\frac{1}{2}+\frac{1}{4}K_{1,2L}\right)M_1. \label{ad2}
\eea
Similarly, for $\bar{\tau}'(0)$ we have
\bea
\bar{\Sigma}_1&=&(-)^{L-1}\sum_{i=1}^{L-1}\bar{M}_{2L}\prod_{j=1}^{i-1}(-2+K_{2L,2j-1})P_{2L,2j}(-2+K_{2L,2i-1})
\\
&\times&\prod_{k=i+1}^{L-1}(-2+K_{2L,2k-1})P_{2L,2k}(-2+K_{2L,2L-1}),\nonumber \\
\bar{\Sigma}_2&=&(-)^{L}\sum_{i=1}^{L-1}\bar{M}_{2L}\prod_{j=1}^{i-1}(-2+K_{2L,2j-1})P_{2L,2j}P_{2L,2i}\\
&\times&\prod_{k=i+1}^{L-1}(-2+K_{2L,2k-1})P_{2L,2k}(-2+K_{2L,2L-1}),\nonumber
\eea
\bea
\bar{\Sigma}_3&=&(-)^{L-1}(-2+K_{2L-2,2L-1})\bar{M}_{2L-2}\prod_{j=1}^{L-2}(-2+K_{2L-2,2j-1})P_{2L-2,2j}(-2+K_{2L-2,2L-3}),\\
\bar{\Sigma}_4&=&(-)^L\bar{M}_{2L}\prod_{j=1}^{L-1}(-2+K_{2L,2j-1})P_{2L,2j}.
\eea
Therefore,
\bea
\bar{\tau}(0)^{-1}\bar{\Sigma}_1&=&-\sum_{i=1}^{L-1} \left(P_{2i,2i+2}-\frac{1}{2}K_{2i,2i+1}K_{2i+1,2i+2}-\frac{1}{2}K_{2i+1,2i+2}K_{2i,2i+1}+\frac{1}{4}K_{2i+1,2i+2}\right) \label{ad3}\\
\bar{\tau}(0)^{-1}\bar{\Sigma}_2&=&\sum_{i=1}^{L-1}\left(-\frac{1}{2}+\frac{1}{4}K_{2i-1,2i}\right)\\
\bar{\tau}(0)^{-1}\bar{\Sigma}_3&=&-P_{2L,2}\left(\bar{M}^{-1}_{2L}\bar{M}_{2}-\frac{1}{2}\bar{M}^{-1}_{2L}\bar{M}_{2} K_{12}-\frac{1}{2}K_{2L,1}\bar{M}^{-1}_{2L}\bar{M}_{2}+\frac{1}{4}K_{2L,1}\bar{M}^{-1}_{2L}\bar{M}_{2} K_{12}\right) \label{spterm1}\\
\bar{\tau}(0)^{-1}\bar{\Sigma}_4&=&-\frac{1}{2}+\frac{1}{4}K_{2L-1,2L} \label{ad4}
\eea
We note that the last term in eq.(\ref{spterm1}) can be simplified as
\bea
&&\left(P_{2L,2}K_{2L,1}\bar{M}_{2L}^{-1}\bar{M}_2K_{12}\right)^{i_1,j_2,j_{2L}}_{j_1,i_2,i_{2L}}\\\no
&=&(P_{2L,2})^{j_{2L},j_2}_{\,e,\,\,\,c}(K_{2L,1})^{a,e}_{j_1,d}(\bar{M}^{-1})^d_{i_{2L}}(\bar{M})^c_b(K_{12})^{i_1,b}_{a,i_2}\\\no
&=&\de^{j_2}_e\de^{j_{2L}}_c\de^a_d\de^e_{j_1}\bar{m}^{-1}_{i_{2L}}\de^d_{i_{2L}}\bar{m}_b\de^c_b\de^b_a\de^{i_1}_{i_2}\\
&=&\de^{i_1}_{i_2}\de^{j_2}_{j_1}\de^{j_{2L}}_{i_{2L}}=(K_{12})^{i_1,j_2}_{j_1,i_2}.
\eea
So it turns out that the nearest neighbor interactions still cancels even for the twisted spin chain. Finally, by adding up eqs.(\ref{ad1})-(\ref{ad2}) and (\ref{ad3})-(\ref{ad4}), we get the Hamiltonian in eq.(\ref{totalH}).
\section{Zero momentum condition}\label{appendixb}
We change the transfer matrices into a form much easier for us to compute by means of permutation operators.
\bea
\tau(0)&=&(-)^L\tr_0 M_0 P_{01}(-2+K_{02})\cdots P_{0,2L-1}(-2+K_{0,2L})\\\no
&=&(-)^L\tr_0 P_{0,2L-1}M_{2L-1} P_{2L-1,1}(-2+K_{2L-1,2})\cdots P_{2L-1,2L-3}(-2+K_{2L-1,2L-2})(-2+K_{0,2L})\\\no
&=&(-)^{L}(-2+K_{2L-1,2L-2})M_{2L-1}\prod_{i=1}^{L-1}P_{2L-2i+1,2L-2i-1}\prod_{j=1}^{L-1}(-2+K_{2j+1,2j}).\\\no
\bar{\tau}(0)&=&(-)^L\tr_{\bar{0}}\bar{M}_{\bar{0}}(-2+K_{\bar{0}1})P_{\bar{0}2}\cdots (-2+K_{\bar{0},2L-1})P_{\bar{0},2L}\\\no
&=&(-)^L(-2+K_{23})P_{24}\cdots (-2+K_{2,2L-1})P_{2,2L}\left[\tr_{\bar{0}}\bar{M}_{\bar{0}}(-2+K_{\bar{0}1})P_{\bar{0}2}\right]\\\no
&=&(-)^L\prod_{i=1}^{L-1}(-2+K_{2i,2i+1})\prod_{j=1}^{L-1}P_{2L-2j,2L-2j+2}\bar{M}_2(-2+K_{21}).
\eea
Therefore after some cancellations, we get
\bea
&&\tau(0)\bar{\tau}(0)\\\no
&=&2^{2(L-1)}\prod_{i=1}^{L-1}P_{2L-2i+1,2L-2i-1}(-2+K_{1,2L})M_1\bar{M}_{2L}(-2+K_{2L,1})\prod_{j=1}^{L-1}P_{2L-2j,2L-2j+2}.
\eea
We can obtain the component of the above operator by acting on a given basis
\bea
&&\left[\tau(0)\bar{\tau}(0)\right]^{I_1,J_2,\cdots I_{2L-1},J_{2L}}_{J_1,I_2,\cdots J_{2L-1},I_{2L}}\\\no
&=&2^{2(L-1)}\left(\prod_{i=1}^{L-1}P_{2L-2i+1,2L-2i-1}\right)^{\,b\,,\,\,I_3,\cdots I_{2L-1}}_{J_1,J_3,\cdots J_{2L-1}}\left[(-2+K_{1,2L})M_1\bar{M}_{2L}(-2+K_{2L,1})\right]^{I_1,J_{2L}}_{\,\,b,\,\,\,a}\\\no
&\times& \left(\prod_{j=1}^{L-1}P_{2L-2j,2L-2j+2}\right)^{J_2,J_4,\cdots J_{2L-2},\,a}_{I_2,I_4,\cdots I_{2L-2},I_{2L}}.
\eea
Since
\bea
\left(\prod_{j=1}^{L-1}P_{2L-2j,2L-2j+2}\right)^{J_2,J_4,\cdots J_{2L-2},\,a}_{I_2,I_4,\cdots I_{2L-2},I_{2L}}&=&\de^a_{I_2}\de^{J_2}_{I_4}\de^{J_4}_{I_6}\cdots \de^{J_{2L-2}}_{I_{2L}},\\
\left(\prod_{i=1}^{L-1}P_{2L-2i+1,2L-2i-1}\right)^{\,b\,,\,\,I_3,\cdots I_{2L-1}}_{J_1,J_3,\cdots J_{2L-1}}&=&\de^{I_3}_{J_1}\de^{I_5}_{J_3}\cdots \de^{I_{2L-1}}_{J_{2L-3}}\de^{\,\,b}_{J_{2L-1}}.
\eea
\bea
\left[(-2+K_{1,2L})M_1\bar{M}_{2L}(-2+K_{2L,1})\right]^{I_1,J_{2L}}_{\,\,b,\,\,\,a}
=2^2m_{I_1}\bar{m}_{J_{2L}}\de^{I_1}_b\de^{J_{2L}}_a,
\eea
we find
\bea
\left[\tau(0)\bar{\tau}(0)\right]^{I_1,J_2,\cdots I_{2L-1},J_{2L}}_{J_1,I_2,\cdots J_{2L-1},I_{2L}}
=2^{2L}\omega^{-ms_{I_1}+ms_{I_2}}\de^{J_{2L}}_{I_2}\de^{J_2}_{I_4}\de^{J_4}_{I_6}\cdots \de^{J_{2L-2}}_{I_{2L}}\cdot \de^{I_3}_{J_1}\de^{I_5}_{J_3}\cdots \de^{I_{2L-1}}_{J_{2L-3}}\de^{I_1}_{J_{2L-1}}.
\eea
So
\bea
&&\left[\tau(0)\bar{\tau}(0)\right]\cdot \tr(\gamma^m Y^{I_1}Y^{\dg}_{I_2}\cdots Y^{I_{2L-1}}Y^{\dg}_{I_{2L}})\\\no
&=&\left[\tau(0)\bar{\tau}(0)\right]^{I_1,J_2,\cdots I_{2L-1},J_{2L}}_{J_1,I_2,\cdots J_{2L-1},I_{2L}}\cdot \tr(\gamma^m Y^{J_1}Y^{\dg}_{J_2}\cdots Y^{J_{2L-1}}Y^{\dg}_{J_{2L}})\\\no
&=&2^{2L}\omega^{-ms_{I_1}+ms_{I_2}}\tr(\gamma^m Y^{I_3}Y^{\dg}_{I_4}\cdots Y^{I_{2L-1}}Y^{\dg}_{I_{2L}}Y^{I_{1}}Y^{\dg}_{I_{2}})\\\no
&=&2^{2L}\omega^{-ms_{I_1}+ms_{I_2}}\tr(Y^{I_{1}}Y^{\dg}_{I_{2}}\gamma^m Y^{I_3}Y^{\dg}_{I_4}\cdots Y^{I_{2L-1}}Y^{\dg}_{I_{2L}})\\\no
&=&2^{2L}\tr(\gamma^m Y^{I_1}Y^{\dg}_{I_2}\cdots Y^{I_{2L-1}}Y^{\dg}_{I_{2L}}).
\eea
which leads to
\bea
\frac{1}{2^{2L}}\tau(0)\bar{\tau}(0)=\mathbb{I}.
\eea
\section{The $osp(6|4)$ algebra}\label{appendixc}
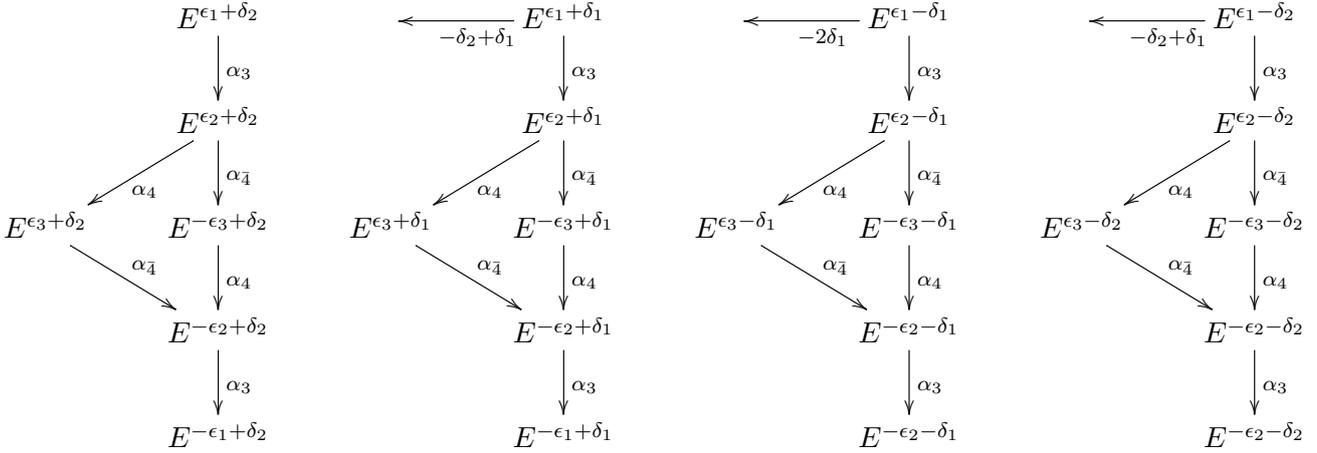
\begin{figure}[H]
\be
\xymatrix{
&E^{\ep_1+\de_2}\ar[d]^{\al_3}&&E^{\ep_1+\de_1}\ar[d]^{\al_3}\ar[l]^{-\de_2+\de_1}&&E^{\ep_1-\de_1}\ar[d]^{\al_3}\ar[l]^{-2\de_1}&&E^{\ep_1-\de_2}\ar[d]^{\al_3}\ar[l]^{-\de_2+\de_1}&\\
&E^{\ep_2+\de_2}\ar[dl]^{\al_4}\ar[d]^{\al_{\bar4}}&&E^{\ep_2+\de_1}\ar[dl]^{\al_4}\ar[d]^{\al_{\bar4}}&&E^{\ep_2-\de_1}\ar[dl]^{\al_4}\ar[d]^{\al_{\bar4}}&&E^{\ep_2-\de_2}\ar[dl]^{\al_4}\ar[d]^{\al_{\bar4}}&\\
E^{\ep_3+\de_2}\ar[dr]^{\al_{\bar4}}&E^{-\ep_3+\de_2}\ar[d]^{\al_4}&E^{\ep_3+\de_1}\ar[dr]^{\al_{\bar4}}&E^{-\ep_3+\de_1}\ar[d]^{\al_4}&E^{\ep_3-\de_1}\ar[dr]^{\al_{\bar4}}& E^{-\ep_3-\de_1}\ar[d]^{\al_4}&E^{\ep_3-\de_2}\ar[dr]^{\al_{\bar4}}&E^{-\ep_3-\de_2}\ar[d]^{\al_4}&\\
&E^{-\ep_2+\de_2}\ar[d]^{\al_3}&&E^{-\ep_2+\de_1}\ar[d]^{\al_3}&&E^{-\ep_2-\de_1}\ar[d]^{\al_3}&&E^{-\ep_2-\de_2}\ar[d]^{\al_3}&\\
&E^{-\ep_1+\de_2}&&E^{-\ep_1+\de_1}&&E^{-\ep_2-\de_1}&&E^{-\ep_2-\de_2}&
}\\\no
\ee
\caption{\footnotesize{The weight diagram for $\mathscr{G}_{\bar{1}}$, where the four vertical weight sub-diagram are the weight diagram $\bm{6}$ of $so(6)$, while themselves are in the $\bm{4}$ of $sp(4)$ which correspond to the horizontal sub-diagram.}}
\label{Odd}
\end{figure}

According to Kac's classification of Lie superalgebra, the $osp(6|4)$ belongs to $D(3,2)$
basic Lie superalgebra,
\bea
\mathscr{G}=osp(6|4),\quad
\mathscr{G}_{\bar{0}}=so(6)\oplus sp(4) ,\quad
\mathscr{G}_{\bar{1}}=(\bm{6},\bm{4}).
\eea
The $\bar{0},\bar{1}$ refer to the $\mathbb{Z}_2$ grading, and the $\bm{6},\bm{4}$ means that the odd part generators $\mathscr{G}_{\bar{1}}$ are in the $\bm{6}$ and $\bm{4}$ representations of the even part $\mathscr{G}_{\bar{0}}$, $\textit{i.e.}$ in the $\bm{6}$ of $so(6)$ and $\bm{4}$ of $sp(4)$. The total 24 odd generators are presented on the Fig.~\ref{Odd}, where we denote $E^{\al}$ as the generators of the algebra, for $\al\in\Delta$.\\
\par The rank of the $osp(6|4)$ algebra is 5, and the root system is
\bea
{\Delta}_{\bar{0}}&=&\{\pm\ep_1\pm\ep_2,\pm\ep_2\pm\ep_3,\pm2\de_1,\pm2\de_2,\pm\de_1\pm\de_2\},\\
{\Delta}_{\bar{1}}&=&\{\pm\ep_1\pm\de_1,\pm\ep_2\pm\de_1,\pm\ep_3\pm\de_1,\pm\ep_1\pm\de_2,\pm\ep_2\pm\de_2,\pm\ep_3\pm\de_2\}.
\eea
where $\de_{1,2},\ep_{1,2,3}$ are two basis satisfy $(\de_i,\de_j)=-\de_{ij}$, $(\ep_i,\ep_j)=\de_{ij}$, $(\de_i,\ep_j)=0$.
The distinguished simple root system is
\bea
\Delta^{0}=\{\de_1-\de_2,\de_2-\ep_1,\ep_1-\ep_2,\ep_2-\ep_3,\ep_2+\ep_3\}.
\eea
we label the simple roots as $\al_1,\al_2,\al_3,\al_4,\al_{\bar4}$ in above giving order. The distinguished simple root system has exactly one odd root, other possible simple root systems can be obtained by odd Weyl reflections.
For our purpose, the \textit{symmetric} Cartan matrix is more useful than the asymmetric definitions  and is defined by,
\be
M_{jj'}=(\al_i,\al_{j'}).
\ee
In the distinguished simple root system, it has the form,
\be
M_{jj'}=\matr{ccccc}{
  -2&         +1 &     &        &         \\
  +1&          &    -1 &        &         \\
    &        -1&    +2 &      -1&      -1 \\
    &          &     -1&      +2&         \\
    &          &     -1&        &      +2
  }.
\ee
\def\g#1{\save
 [].[dr]!C="g#1"*[F]\frm{}\restore}%
\subsection{Odd Weyl reflections}
We know that Dynkin diagram is not unique for simple Lie superalgebra. We extend the ordinary Weyl reflections (reflections with respect to even roots),
\be
w_\alpha\beta=\beta-2\frac{(\beta,\alpha)}{(\alpha,\alpha)}\alpha,
\ee
for $\beta\in\Delta, \alpha\in\Delta_0$, to include the case with respect to odd roots as well,
\be
\begin{aligned}
&w_\alpha\beta=\beta-2\frac{(\beta,\alpha)}{(\alpha,\alpha)}\alpha,&\text{if}\;(\alpha,\alpha)\neq0,\\
&w_\alpha\beta=\beta+\alpha, \qquad\qquad&\text{if}\;(\alpha,\alpha)=0\;\text{and}\;(\alpha,\beta)\neq0,\\
&w_\alpha\beta=\beta, \qquad\qquad&\text{if}\;(\alpha,\alpha)=0\;,\;(\alpha,\beta)=0\;\text{and}\;\beta\neq\alpha,\\
&w_\alpha\alpha=-\alpha.
\end{aligned}
\ee
Begin with the distinguished simple root system, using the odd root Weyl reflections upon each root, we get a new simple root system and this procedure goes on and on. Here we give some examples. The distinguished simple root system of $osp(6|4)$ is
\bea
\Delta^{0}=\{\de_1-\de_2,\de_2-\ep_1,\ep_1-\ep_2,\ep_2-\ep_3,\ep_2+\ep_3\}.
\eea
\begin{figure}[H]
  \centering
  \includegraphics[width=5cm]{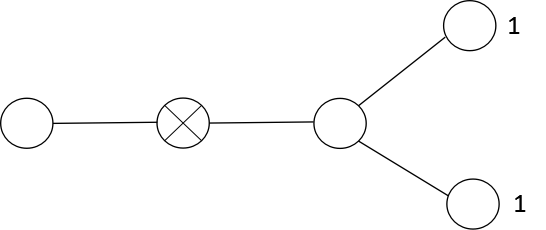}
\end{figure}
Applying the Weyl reflection with respect to the second simple root, we get
\be
w_{\de_2-\ep_1}(\Delta^0)=\{\de_1-\ep_1,-\de_2+\ep_1,\de_2-\ep_2,\ep_2-\ep_3,\ep_2+\ep_3\}.
\ee
\begin{figure}[H]
  \centering
  \includegraphics[width=5cm]{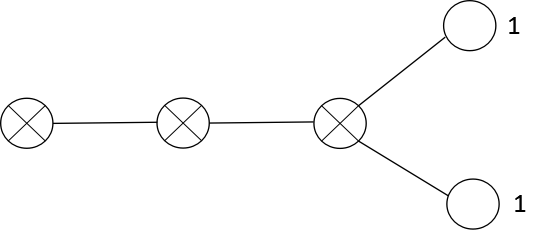}
\end{figure}
Other examples are
\be
w_{\de_1-\ep_1}(w_{\de_2-\ep_1}(\Delta^0))=\{-\de_1+\ep_1,\de_1-\de_2,\de_2-\ep_2,\ep_2-\ep_3,\ep_2+\ep_3\},
\ee
\begin{figure}[H]
  \centering
  \includegraphics[width=5cm]{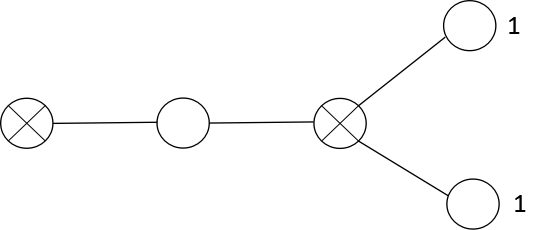}
\end{figure}
\be
w_{\de_2-\ep_2}(w_{\de_2-\ep_1}(\Delta^0))=\{\de_1-\ep_1,\ep_1-\ep_2,-\de_2+\ep_2,\de_2-\ep_3,\de_2+\ep_3\},
\ee
\begin{figure}[H]
  \centering
  \includegraphics[width=5cm]{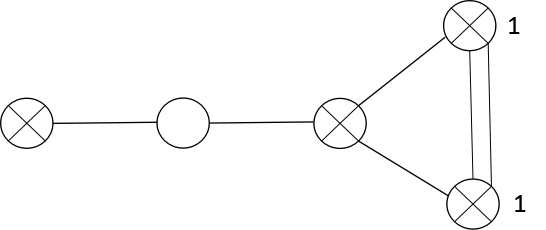}
\end{figure}
\be
w_{\de_2-\ep_2}(w_{\de_1-\ep_1}(w_{\de_2-\ep_1}(\Delta^0)))=\{-\de_1+\ep_1,\de_1-\ep_2,-\de_2+\ep_2,\de_2-\ep_3,\de_2+\ep_3\},
\ee
\begin{figure}[H]
  \centering
  \includegraphics[width=5cm]{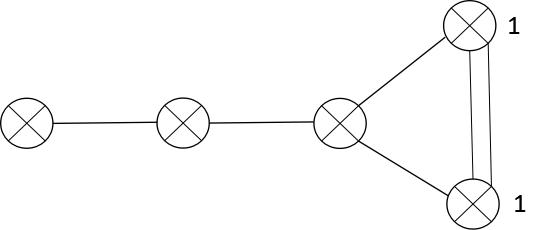}
\end{figure}
\be
w_{\de_2+\ep_3}(w_{\de_2-\ep_2}(w_{\de_2-\ep_1}(\Delta^0)))=\{\de_1-\ep_1,\ep_1-\ep_2,\ep_3+\ep_2,-\de_2-\ep_3,2\de_2\}.
\ee
\begin{figure}[H]
  \centering
  \includegraphics[width=6.5cm]{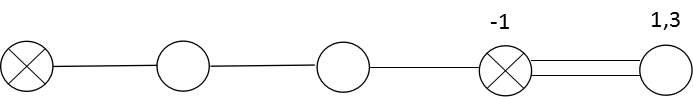}
\end{figure}

\end{appendix}

\end{document}